\documentclass[jgrga]{agutex}
\usepackage{graphicx}        
\usepackage{color}           
\usepackage{url}             

\newcommand{\BE}{\begin{equation}}
\newcommand{\EE}{\end{equation}}
\newcommand{\BA}{\begin{eqnarray}}
\newcommand{\EA}{\end{eqnarray}}
 \newcommand{\fig}[1]{Figure~\ref{fig:#1}}
 \newcommand{\figsss}[1]{Figures~\ref{fig:#1}}

 \newcommand{\sect}[1]{Section~\ref{sect:#1}}
 \newcommand{\sectss}[2]{Sections~\ref{sect:#1}-\ref{sect:#2}}
 \newcommand{\eq}[1]{Equation~(\ref{eq:#1})}
 \newcommand{\eqs}[2]{Equations~(\ref{eq:#1}) and (\ref{eq:#2})}

\newcommand{\rmd}{ {\mathrm d} }
\renewcommand{\vec}[1]{ {\bf #1} }
\newcommand{\uvec}[1]{ \hat{\bf #1} }

\newcommand{\eg}{\textit{e.g.}}

\newcommand{\ie}{\textit{i.e.}}
\newcommand{\insitu}{\textit{in situ}}

\newcommand{\aB}{<\!\!B\!\!>}

\newcommand{\Bo}{B_0}  
\newcommand{\cP}{c_{\rm P}}  
\newcommand{\cS}{c_{\rm S}}  
\newcommand{\degree}{^\circ} 

\newcommand{\iA}{i}  
\newcommand{\kms}{km.s$^{-1}$} 
\newcommand{\lA}{\lambda}

\newcommand{\pA}{\phi}
\newcommand{\phimax}{\varphi_{\rm max}}
\newcommand{\pobs}{\mathcal{P}_{\rm obs}}
\newcommand{\pobsl}{\mathcal{P}_{\rm obs}(\lA)}
\newcommand{\pobsAl}{\mathcal{P}_{\rm obs}(|\lA|)}
\newcommand{\pobsAi}{\mathcal{P}_{\rm obs}(|\iA|)}

\newcommand{\pvphi}{\mathcal{P}_{\varphi}}
\newcommand{\Ro}{R}  
\newcommand{\tA}{\theta}
\newcommand{\ur}{\hat{\bf u}_{\rho}}
\newcommand{\up}{\hat{\bf u}_{\varphi}}

\renewcommand{\aap}{{\it Astron. Astrophys.}}

\renewcommand{\apjl}{{\it Astrophys. J. Lett.}}
\renewcommand{\apjs}{{\it Astrophys. J. Supp. Series}}

\authorrunninghead{JANVIER ET AL.}

\titlerunninghead{Properties of Flux Ropes in the Solar Wind}

\authoraddr{Corresponding author: M. Janvier,
Department of Mathematics, University of Dundee, Dundee DD1 4HN, Scotland,  
       United Kingdom (mjanvier at maths.dundee.ac.uk)}

\begin{document}



\title{{In Situ} Properties of Small and Large Flux Ropes in the Solar Wind \\
}

\authors{M. Janvier,\altaffilmark{1}
P. D\'emoulin,\altaffilmark{2} S. Dasso,\altaffilmark{3,4,5}}

\altaffiltext{1}{Department of Mathematics, University of Dundee, Dundee DD1 4HN, Scotland, 
        United Kingdom}

\altaffiltext{2}{Observatoire de Paris, LESIA, UMR 8109 (CNRS), 92195 Meudon Principal Cedex, France}

\altaffiltext{3}{Instituto de Astronom\'\i a y F\'\i sica del Espacio, UBA-CONICET, CC. 67, Suc. 28,
       1428 Buenos Aires, Argentina}

\altaffiltext{4}{Departamento de F\'\i sica, Facultad de Ciencias Exactas y Naturales, 
       Universidad de Buenos Aires, 1428 Buenos Aires, Argentina}

\altaffiltext{5}{Departamento de Ciencias de la Atm\'osfera y los Oc\'eanos, Facultad de Ciencias Exactas y Naturales, Universidad de Buenos Aires, 1428 Buenos Aires, Argentina}


\begin{abstract}
  Two populations of twisted magnetic field tubes, or flux ropes ({hereafter,} FRs), are detected by \insitu\ measurements in the solar wind.
While small FRs are crossed by the observing spacecraft within few hours, {with a radius typically less than 0.1AU}, larger FRs, or magnetic clouds ({hereafter,} MCs), have durations of about half a day.  
  The main aim of this study is to compare the properties of both populations of FRs observed by the \textit{Wind} spacecraft at 1~AU.
  To do so, we use standard correlation techniques for the FR parameters, as well as histograms and more refined statistical methods. 
Although several properties seem at first different for small FRs and MCs, we show that they are actually {governed by the same propagation physics}.
For example, we observe no \insitu\ signatures of expansion for small FRs, contrary to MCs. We demonstrate that this result is in fact expected: 
small FRs expand similarly to MCs, as a consequence of a total pressure balance with the surrounding medium, but the expansion signature is well hidden by {velocity} fluctuations.
   Next, we find that the FR radius, velocity and magnetic field strength are all positively correlated{, with correlation factors than can reach a value $>$0.5}. 
This result indicates a remnant trace of the FR ejection process from the corona. 
We also find a larger FR radius at the apex than at the legs {(up to three times larger at the apex)}, for FR observed at 1 AU. 
   Finally, assuming that the detected FRs have a large-scale configuration in the heliosphere, we derived the mean axis shape from the probability distribution of the axis orientation.  
We therefore interpret the small FR and MC properties in a common framework of FRs interacting with the solar wind, and we disentangle the physics present behind their common and different features.
\end{abstract}



\begin{article}

\section{Introduction}
     \label{sect:Introduction} 

Twisted magnetic flux tubes, or flux ropes (FRs), are at least present in all astrophysical magnetic media that can be analysed with enough spatial resolution {\citep[\textit{e.g.},][]{Dasso09b,Fan09,Linton09}}. These range from the convective zone and the atmosphere of the Sun, to the solar wind (SW) and planet magnetospheres. 
Some FRs of solar origin are expected to be formed at the base of the convective zone, in the solar interior, and rise through it all the way {to emerge across the photosphere and to form} 
active regions as modelled by numerical simulations \citep[\textit{e.g.},][]{Fan09,Pinto13,Weber13}. Good observational evidences are given by local helioseismology, which begins to detect the FR arrival below the photosphere \cite[\textit{e.g.},][]{Toriumi13}, but also with photospheric magnetograms, which indicate the presence of a twisted magnetic field during an active region emergence \cite[\textit{e.g.},][]{Luoni11}. Higher up in the solar atmosphere, evidences of twisted magnetic field have been found around filaments and in coronal mass ejections (CMEs, \eg\ \citet{Aulanier10}; \citet{Cheng13}; see the review of \citet{Schmieder13}).   FRs are also detected both at the magnetopause of the Earth magnetosphere as flux transfer events (\eg\ \citet{Eastwood12}; \citet{Zhang12}) as well as in the magnetotail (see the review of \citet{Linton09}). \citet{Cartwright10} proposed that some FRs detected in the interplanetary medium could be formed in the heliospheric current sheet (HCS).

 Coherent magnetic field rotations are measured \insitu\ in the SW on a time scale of {half an hour to} hours, and are interpreted as small FRs  \citep{Moldwin00,Cartwright08,Feng07}. {The small FRs are identified by the coherent rotation of the magnetic field and its enhanced strength compared to the surrounding SW.
After performing a fit with a flux rope model (see below), these boundaries are refined in the FR frame where axial and azimuthal components are separated (see \citet{Feng06} for further informations).}  A larger class of events, called small transients, has been recently identified with a definition emphasizing the plasma characteristics \citep{Yu14}. 

 On a time scale of half a day, larger FRs, also called magnetic clouds (MCs), are regularly detected by spacecraft \citep[][and references therein]{Bothmer98,Lynch05,Lepping10}.  {They are defined, similarly as small FRs, by a coherent rotation and an enhanced magnetic field strength with the additional condition of a proton temperature lower than the typical temperature found in the SW travelling at the same speed (see \citet{Lepping03b} for further informations).  As for small FRs, the definition of MC boundaries could be ambiguous, and different authors may not agree \cite[\textit{e.g.},][]{Dasso06,Al-Haddad13}.}  
 
{ Previous studies have not associated small FRs to the crossing of MCs near their boundaries (\ie\ at large impact parameters). Indeed, a size of a factor ten smaller would require a spacecraft crossing very close to the MC boundary and a large rotation of the magnetic field would not be observed \cite[see Figure~5 of ][]{Demoulin13}.  It was also argued that small FRs and MCs are formed by a different process \cite[\textit{e.g.},][]{Cartwright08}. However, other authors \cite[\textit{e.g.},][]{Feng07} have also pointed toward a common one. Since} a small fraction of magnetic clouds have been found with a duration of several hours, making the distinction between small FRs and MCs is difficult{, as solely based on the duration of the events.}  As such, these two populations of FRs slightly overlap in radius \citep{Janvier14}. In the following, both FR populations are simply referred to as FRs, while we refer to small FRs and MCs to distinguish the two populations of detected FRs.
 
 Small FRs and MCs have several common, or at least similar, properties.
\begin{itemize}
  \item At 1~AU, they all travel away from the Sun with speeds mostly in the range of the slow SW (300\,--\,600 \kms), with the exception of a small fraction of MCs travelling significantly faster (up to $\sim$ 1000 \kms).
  \item Both small FRs and MCs are relatively well fitted by the simple Lundquist FR model \citep{Lundquist50}.  An output of this fit is the orientation of the FR axis.
The distribution of the axis longitude is almost uniform for both FR groups. A uniform distribution is also found for the axis latitude for MCs, while small FRs tend to be more parallel to the ecliptic \citep{Feng08,Janvier13}. 
  \item The plasma density $N_p$ is nearly uniform within small FRs, as shown in Figure~8 of \citet{Cartwright10}.  In MCs, the $N_p$ distribution is also approximatively uniform within the structure, with comparable values as for small FRs.  Still, there is a weak tendency of increasing $N_p$ values at the MC rear.  This tendency is however variable within the considered MC sets, as shown with the large dispersion of $N_p$ at the MC rear \citep[see Figure~3 of][]{Lepping03b}.
  \item Small FRs and MCs have typically a sheath of accumulated SW plasma and magnetic field at their front, although this characteristic is much weaker for small FRs and only well seen in superposed time interval analysis (\citep[see Figure~8 of][]{Cartwright10}.
  \item {Both FRs and MCs present signatures of magnetic reconnection due to their propagation in the interplanetary medium and solar wind interaction \citep{Tian10, Lavraud14}. This indicates similar propagation properties.}
\end{itemize}

Then, apart from their sizes, small FRs and MCs also have different properties:
\begin{itemize}
  \item The field strength $B$ is typically enhanced in MCs compared with small FRs. 
  \item MCs have a proton temperature $T_p$ lower than in the typical SW travelling at the same speed (this is part of the MC definition), while no significant and typical temperature variations are observed across small FRs \citep{Feng07}.    
  \item The radius $R$ of small FRs statistically increases with the distance $D$ from the Sun as: $R \propto D^{0.43}$ \citep{Cartwright10}.  This expansion is much smaller than for MCs ($R \propto D^{0.6 \rm{-} 1}$, see  {\citet{Bothmer94};} \citet{Gulisano12} and references therein).  Also, contrary to MCs, no expansion is directly detected by \insitu\ measurements for small FRs, and more generally for small transients \citep{Moldwin00,Feng07,Cartwright08,Kilpua12,Yu14}.
  \item While $T_p$ is larger and $B$ is lower in small FRs in comparison with MCs, a similar $N_p$ value implies a larger plasma $\beta$ ($\approx 1$) in small FRs {\citep{Cartwright10}}, while $\beta <0.1$ in MCs   {\citep[\eg ][]{Lepping03b}.  As such, the magnetic field clearly dominates in MCs, while due to weak temperature and plasma density variations across small FRs, the gradient of the plasma pressure may be balanced by the magnetic tension insuring the FR coherence.  However, the stability of small FRs over a distance of 1 AU cannot be tested with present observations}.
  \item The distribution of the FR duration is double-peaked \citep{Cartwright08}. This was confirmed by analysing the distribution of the FR radius in \citet{Janvier14}, which showed that small FRs have a steep power law distribution, while MCs have a Gaussian-like distribution.  
\end{itemize}
    
The main aim of this study is to further analyse the small FRs properties and to compare them with that of MCs. The statistical distributions of the number of detected small FRs and MCs, as well as their different characteristics, indicate a different origin for these two populations. {Are those differences due to different mechanisms occuring in the solar corona, or are small FRs created directly in the heliosphere? Since} all are twisted magnetic field structures evolving in the SW, they also share common physical mechanisms that tend to create comparable characteristics. In order to disentangle their specific origin properties from their common propagation properties, we propose to better characterise both small FRs and MCs populations by using the same statistical analysis tools.

We first revisit the expansion properties of small FRs and MCs in \sect{Expansion}. In particular, we analyse the origin of this expansion, {and the reason why MCs are the only FRs for which an expansion is detected} by \insitu\ velocity measurements. Then, we summarise the properties of the data sets analysed, as well as their modelling in \sect{Observations}.  In \sect{Correlations}, we analyse the correlations between the FRs parameters in order to quantitatively compare small FR and MC properties.  In \sect{Orientation}, we investigate more specifically the distributions of the FR axis, so as to deduce the mean shape of the axis.  Finally, in \sect{Conclusion}, we summarise our results in the context of previously known results and conclude on small FR versus MC properties.

\section{Flux Rope Expansion} 
      \label{sect:Expansion} 

\subsection{Size Expansion with Solar Distance } 
      \label{sect:Expansion-Size}
 
Occurrences of two spacecraft observing the same FR at different solar distances $D$ are very rare. As such, the FR expansion is difficult to estimate from the evolution of the same observed FR ({\citet{Bothmer98} followed the evolution of some MCs from Helios to Voyager spacecraft, while the MC carefully studied by \citet{Nakwacki11} at both 1 and 5.4 AU is one of the few exceptional cases}). Therefore, statistical analyses were realised with several spacecraft located at significantly different helio-distances (0.3 to 5 AU for the broadest range) so as to deduce the mean expansion of FRs.
Then, for MCs, \citet{Kumar96} found $R \propto D$, while {\citet{Bothmer94,Bothmer98}} found $R \propto D^{0.8}$ and \citet{Leitner07} found $R \propto D^{0.6}$. 
MCs are therefore expanding as $R \propto D^{0.8 \pm 0.2}$.
Applying the same method, \citet{Cartwright10} found $R \propto D^{0.43}$ for small FRs.  Do small FRs really expand with a factor two less in the power-law exponent than MCs? In the following, we check the coherence of this result and propose an interpretation. 
 
   { A possible origin of the FR expansion with the solar distance could be a gradient of speed between the leading and the trailing edges of CMEs in the corona, as observed with coronagraphs.
However, as shown in numerical simulations, plasma and magnetic forces can modify the coronal velocities.  For example, the inserted flux rope in \citet{Xiong06} is initially out of equilibrium, which implies a fast expansion at the early stage of the flux rope evolution (\eg\ see their Figure~7). Later on, when an approximative force balance is achieved, the expansion rate decreases.  Then, the initial coronal expansion is expected to be modified during the propagation of FRs.}  

 The magnetic field strength in small FRs is found to decrease as $B \propto D^{-0.94}$ \citep{Cartwright10}.  With a cylindrical symmetry hypothesis, this result implies that the axial magnetic flux behaves as $B~R^2 \propto D^{-0.08}$. Therefore, there can only be a weak erosion of the flux with distance between 0.3 and 5.5 AU. For MCs, the magnetic field strength is found to decrease between $D^{-1.8}$ and $D^{-1.3}$ \citep{Kumar96,Leitner07}. Within the large statistical fluctuations of events, this also implies an approximate conservation of magnetic flux with solar distance (with $B~R^2 \propto D^{\alpha}, |\alpha|<0.2$).  The low expansion rate of small FRs, compared to that of MCs, is then coherent with the independent magnetic field measurements and with the approximative verification of the conservation of magnetic flux.

  Furthermore, \citet{Cartwright10} studied the ratio of small FRs magnetic field strength to that of the nearby related SW.  This ratio has a weak dependence with distance ($\propto D^{-0.07}$), and a superposed time interval analysis (see their Figure~8) only shows a weak enhancement of $B$ in small FRs compared to the surrounding SW.   They also found a nearly balanced total pressure (magnetic and thermal) between the inside and the outside of small FRs.  Such a balance of total pressure was shown to be at the origin of the MC expansion \citep{Demoulin09,Gulisano12}. Then, could this balance of pressure also be at the origin of the small FR expansion?
 
 Since there is no significant contrast in density and temperature (at least for protons) between the inside and the outside of the twisted magnetic structure, the total pressure balance almost reduces to a magnetic pressure balance between the inside of the FR and the surrounding SW.  Next, small FRs are typically found nearby the HCS (see Figure 9 of \citet{Cartwright10}), where the radial magnetic field vanishes. There, the SW magnetic pressure is expected to be dominated by the azimuthal component $B_{\varphi}$, which decreases as $\approx 1/D$, while the radial component decrease rate is closer to $\approx 1/D^2$ (e.g., \citet{Mariani90}).  This $B_{\varphi}$ dependence on the solar distance is comparable to the field strength dependence found for small FRs, $B \propto D^{-0.94}$, and in agreement with a balance of magnetic pressure. Imposing flux conservation implies $R \propto D^{0.5}$, which is close to the $R \propto D^{0.43}$ relation found by \citet{Cartwright10}. 
    
 We conclude that the approximative balance of total pressure between the inner and outer sides of FRs is the origin of the common expansion phenomenon for both small FRs and MCs, with the following differences. 
The internal plasma pressure in MCs is negligible with respect to the internal magnetic pressure, and the ambient 
{total SW pressure is decreasing as $\approx D^{-2.8}$, implying a FR radius increasing as $\approx D^{0.8}$ \citep{Demoulin09}. For small FRs, the plasma pressure difference is weak between the inside and the outside, and}
the decrease of the ambient magnetic field (nearby the HCS) follows a $1/D$ law (see previous paragraphs). {Then, the balance of pressure implies a FR radius increasing only as $\approx D^{0.5}$. The different} pressure balance {conditions with the surrounding SW for small FRs and MCs therefore} explain the difference in the observed expansion rates.

\begin{figure}  
 \centerline{ \includegraphics[width=0.5\textwidth]{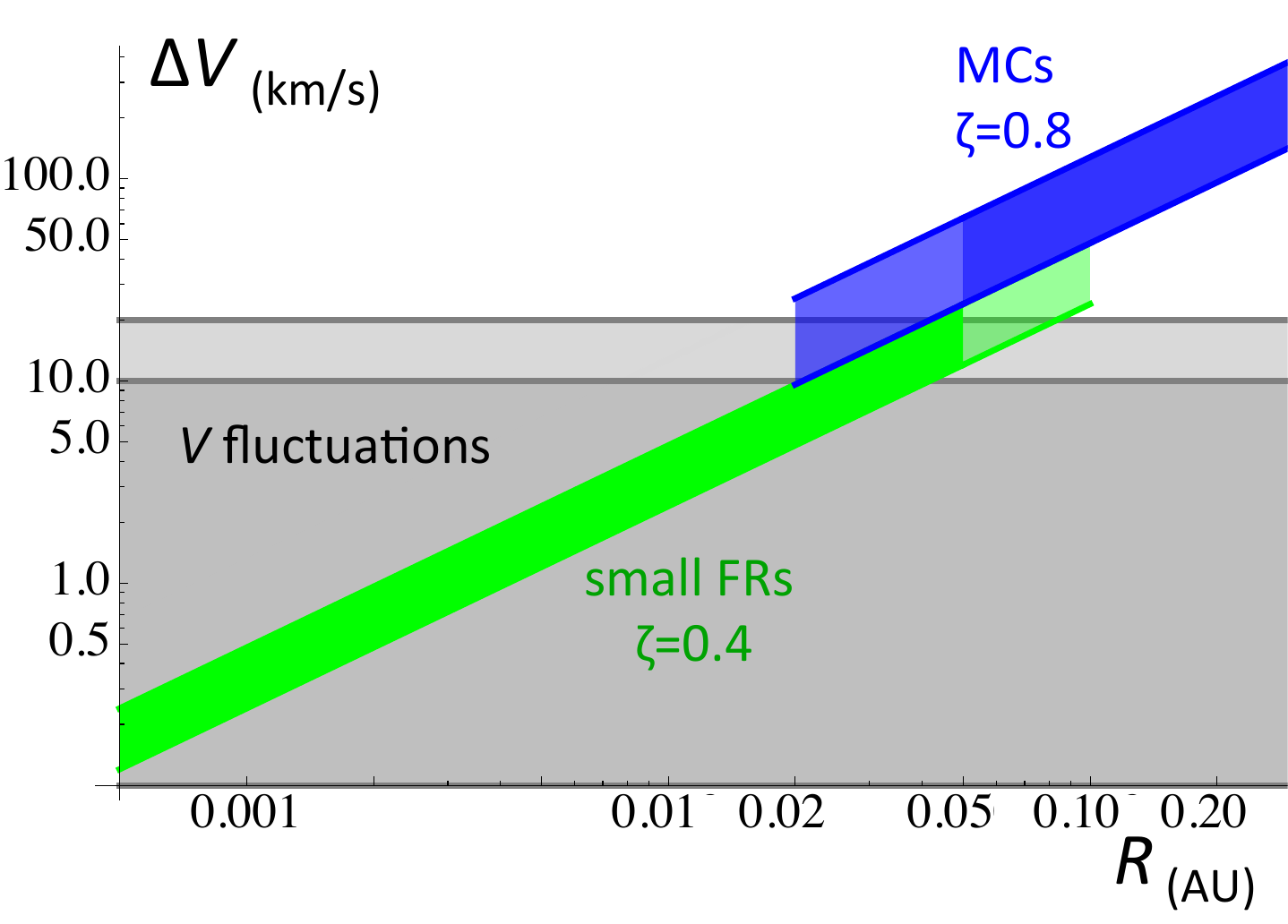} }
\caption{Log-log plot showing the expected expansion velocity $\Delta V$ (in \kms), estimated from \eq{DV}, in function of the FR radius R (in AU). 
The green and blue bands are the expected $\Delta V$ for small FRs and MCs, respectively, estimated from the non-dimensional expansion rate $\zeta$ defined with {statistical studies (\sect{Expansion-Size})}  
The bands have a lighter colour for $R>0.05$ (resp. $R<0.05$) for small FRs (resp. MCs) to indicate the presence of only a small fraction of overlapping cases (see \sect{Observations} and \fig{V(R)_Bo(R)}).  
The vertical extension of the bands is defined by the FR velocity, which is in the range $[300,600]$ \kms\ for small FRs and in the range $[300,800]$ \kms\ for MCs, as typically observed at 1~AU \citep[see \fig{V(R)_Bo(R)}c,d;][]{Feng08, Lepping10}.
The grey bands show the region dominated by the observed velocity fluctuations, with two levels of fluctuations (10 and 20 \kms ). Then, $\Delta V$ is only detected above the upper grey band.
}
 \label{fig:DV}
\end{figure}  

\subsection{\textit{In Situ} Expansion Velocity } 
      \label{sect:Expansion-Velocity}

The radial velocity (away from the Sun) in MCs is generally found to be faster at its front than at its rear, with a typical velocity profile almost linear in time, as long as they are not significantly perturbed (\eg, by an overtaking fast SW stream or another MC, see \citet{Gulisano12} and references therein).  Then, the expansion velocity $\Delta V$ is defined by the difference between the radial velocity at the MC front and at its rear \citep[\eg\,][]{Gosling94,Kilpua13}.  The magnitude of $\Delta V$ is frequently used to characterise how fast a MC, and more generally a FR, expands. However, in order to compare FRs with different sizes, using $\Delta V$ only is misleading since smaller FRs necessarily have smaller $\Delta V$, as follows.  {As a simple example to illustrate our point, we} consider an expanding flux tube of radius $R$ with an expansion velocity $\Delta V$ as measured by a spacecraft orthogonally crossing its axis. 
Then, we also consider another FR with the same velocity pattern but limited to, say, the core of the previous FR (\ie, with a radius to $r<R$).
Its expansion velocity is $\Delta V \; r/R$, since the $V$ profile is linear.  This expansion velocity is small, or even not observable, if the ratio $r/R$ is too small. We conclude that $\Delta V$ is indeed size-dependent and cannot be used alone to characterise the \textit{intrinsic} expansion rate of a FR.     
 
Following the work of \citet{Demoulin09}, we define the non-dimen\-sional expansion rate as
   \begin{equation}   \label{eq:zeta}
   \zeta = \frac{\Delta V}{\Delta t} \frac{D}{V_c^2} \,,
   \end{equation}
where $\Delta t$ is the spacecraft crossing time, $D$ the solar distance and $V_c$ the translation velocity of the FR (more precisely, of its centre).  Following the FR motion away from the Sun, if $\zeta$ is independent of $D$, \eq{zeta} implies a power law for the FR radius as $R \propto D^{\zeta}$ \citep[see Section 3.3 of][]{Gulisano10}, \ie, the same dependence as found from the statistical analysis (\sect{Expansion-Size}).
From \insitu\ observations of the velocity profile of MCs, previous investigations found that $\zeta \approx 0.8$ to $1$ \citep{Demoulin08,Gulisano10,Gulisano12}.  Moreover, since there is almost no dependence of $\zeta$ on $D$ with observations from $D=0.3$ to $5$~AU for non-perturbed MCs, $R \propto D^{0.9 \pm 0.1}$.  This confirms, with different types of data and analysis methods, the statistical results of $R(D)$ for MCs as summarized in \sect{Expansion-Size}.   

Then, with $\zeta$ known, \eq{zeta} can be inverted to predict the expected expansion velocity $\Delta V$ as  
   \begin{equation}   \label{eq:DV}
   \Delta V \approx 2 ~\zeta \frac{R}{D} ~V_c \,,
   \end{equation}
where we suppose a crossing near the FR axis (so a small impact parameter) and almost orthogonal to the axis. 
The estimated $\Delta V$ is shown in \fig{DV} for the observed range of $V_c$ and the typical $\zeta$ values quoted above.  $\Delta V$ is measured along the radial direction, away from the Sun.
For a finite impact parameter and/or a non-orthogonal crossing, $2 R$ in \eq{DV} is replaced by the length of the crossing. It implies that $\Delta V$ decreases for a larger impact parameter {and/or a non-orthogonal crossing.} Next, $\zeta$ is a function of the expansion coefficients in the three principal directions \citep{Demoulin08}. For MCs, the expansion is nearly isotropic \citep{Nakwacki11}, so $\zeta$ is approximately independent of the crossing angle. For small FRs, the axial expansion is unknown. All in all, this implies that \fig{DV} shows the upper limits for the expansion velocity $\Delta V$ for a nearly orthogonal crossing, while it could be a low estimation for crossings more parallel to the FR axis.

The typical velocity fluctuation level observed in MCs is $\approx 20$ \kms\ \citep{Lepping03b}. We perform a similar analysis to compute the level of velocity fluctuations for small FRs, and we obtain a value around $10$ \kms. 
The level of fluctuations in MCs and small FRs are both similar to that found for the typical solar wind, measured with intervals of 10-20 hours and 1-2 hours (taking a larger size interval in the solar wind typically provides a larger level of fluctuations for scales between 1-20 hours).
With these levels of fluctuations, \fig{DV} shows that most of MCs have an expansion velocity well above the velocity fluctuations while the expansion of most small FRs is masked by the velocity fluctuations.
More precisely, there is only 6\% of the small FRs analysed by \citet{Feng08} with $R > 0.05$~AU
(\fig{V(R)_Bo(R)}a,c), and only a small fraction of the smaller ones have $\Delta V$ above the typical fluctuation magnitude. {Moreover, because of their lower size and field strength, small FRs are even more affected by the SW environment, such as an overtaking stream, than MCs.}  It implies that only a {small fraction} 
of small FRs may have an expansion $\Delta V$ just large enough to be detected directly from the \insitu\ velocity measurements (\fig{DV}) {in agreement with the absence of expansion detected locally in small FRs \citep{Moldwin00,Feng07,Cartwright08,Kilpua12,Yu14}.} For nearly all small FRs, the velocity profile across the FR is expected to be nearly constant (apart from the fluctuations and the interactions with the surrounding SW), from which it is incorrect to conclude that they are not expanding since the expected expansion is masked by the velocity fluctuations and the perturbations from the surrounding SW.

\section{Analysed Data Sets} 
      \label{sect:Observations} 

We summarise in this section the general information on the data sets used to further study the properties of the small FRs and MCs.

 We study the population of small FRs from the results of \citet{Feng08}. They identified 125 small FRs between March 1995 and November 2005 using the plasma and magnetic field data from the {\it Wind} spacecraft orbiting near Earth. These results have been extended from \citet{Feng07} to a larger range of time. The authors limited the duration of the coherent rotation of the magnetic field between half and eleven hours, and therefore purposely excluded most of the MCs. In the following, only seven out of the whole set are in the MC size range, indicating that, at most, only a few MCs could remain in the small FR list.    
 
We compare below the properties of these small FRs to those of a population of MCs as analysed by \citet{Lepping10}.  We use a slightly more extended list (Table~2 at \url{http://wind.nasa.gov/mfi/mag_cloud_S1.html}), which, at the date of 24 October 2013, contains the parameters obtained for 121 MCs observed nearby Earth by the {\it Wind} spacecraft from February 1995 to December 2009. However, not all the properties of these MCs are well defined for this set of data. As such, we removed 14 doubtful cases from the above list, for which the handedness could not be determined (flag f in the list), or for which the fitting convergence could not be achieved (flag F), or the estimated spacecraft trajectory was too far from the FR axis (which is determined by the Lundquist fit as the distance where the axial field reverses).  Removing these suspicious cases, all with significant deviations from the Lundquist's model, 107 MCs were left. For all the remaining cases, the spacecraft crossing duration ranges between five and sixty hours, so about a factor ten larger than the average time duration for small FR crossings.
  
 The obvious main difference between small FRs and MCs is their radius $\Ro$ {(see below for its estimation method). This difference is} shown schematically in \fig{DV} by the extensions of $\log \Ro$ for both types. The overlapping region has only few cases because the small-FR distribution is a steep decreasing power-law of $\Ro$ and the distribution of the MCs is a Gaussian-like distribution centered around $R \approx 0.12$~AU with a small standard deviation of $\approx 0.05$~AU \citep{Janvier14}. It implies that only seven small FRs (6\%) have $\Ro >0.05$~AU while only ten MCs (9\%) have $\Ro <0.05$~AU (see these cases in \fig{V(R)_Bo(R)}). Small FR and MC distributions only slightly overlap in radius and $\Ro =0.05$~AU defines a typical size limit between the two FR populations.   {These results indicate that small FRs and MCs are well separated into two lists, with at most a small percentage of incorrect identification.}  

The \insitu\ data only provide a 1D cut of the magnetic field configuration along the spacecraft trajectory. These data are then fitted by a FR model so as to estimate the global properties of the crossed FR, such as its radius $\Ro$ and its axial field strength $\Bo$.  The fitting model used for both datasets is the classical Lundquist's field model, which corresponds to a linear force-free magnetic field with a circular section and a straight axis \citep{Lundquist50}.  This provides an estimation of $\Ro$ and $\Bo$,
as well as the longitude $\pA$ and the latitude $\tA$ of the FR axis, in the Geocentric Solar Ecliptic (GSE) system of reference. 

   {The quality of the fit is characterized by the square root of the chi-square function $\chi$.  Reducing the sample of MCs to MCs with the best fit (or best quality) does not significantly affect the distribution of the fitted parameters such as the axis directions \citep{Janvier13}.   Including the observed expansion rate in the fitted model only has a weak effect for MCs with a radius larger than 0.1~AU \citep{Demoulin08,Nakwacki08}. Since FRs are travelling at different speeds than the surrounding SW, their cross-section is deformed.  However, their cross section is flatten in the radial direction only by a factor 2 to 3 in average \citep{Demoulin13}.  The fit results are more biased as the spacecraft crosses the FR further away from the apex since the curvature of the axis, not included in the Lundquist's model, affects more the magnetic profiles \citep{Owens12}.  Finally, the most critical issue for the derived parameters is the selection of the FR boundaries \citep{Al-Haddad13}.  Such boundaries are better determined in the FR frame \citep{Dasso06,Feng06}.  Indeed, when the boundaries are well selected, the minimum variance and Lundquist's fit techniques have shown to provide close directions \citep{Ruffenach12}.   In summary, a number of tests have been performed previously to check the performances of the Lundquist's fit.   While being the simplest model, none of more complex models (non-linear force-free, non force-free, non circular) have shown to perform better on a large ($>20$) sample of MCs.  Moreover its broad use provides a common method to analyze small FRs and MCs over large data sets.   It would be worth to analyze such large data sets with other methods, but this is outside the scope of this paper. 
}

\begin{figure}  
 \centerline{
     \includegraphics[width=0.19\textwidth]{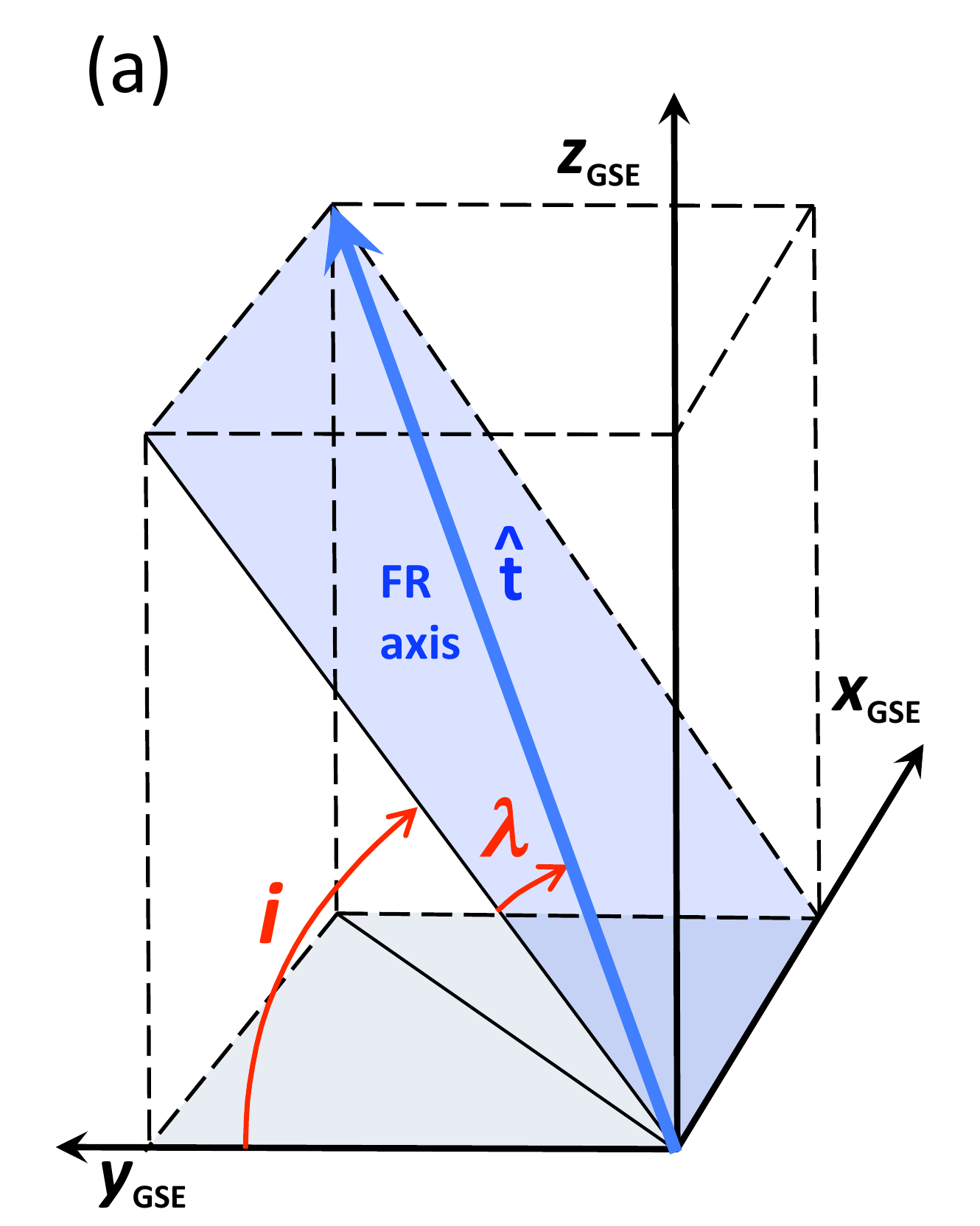}
   \includegraphics[width=0.31\textwidth]{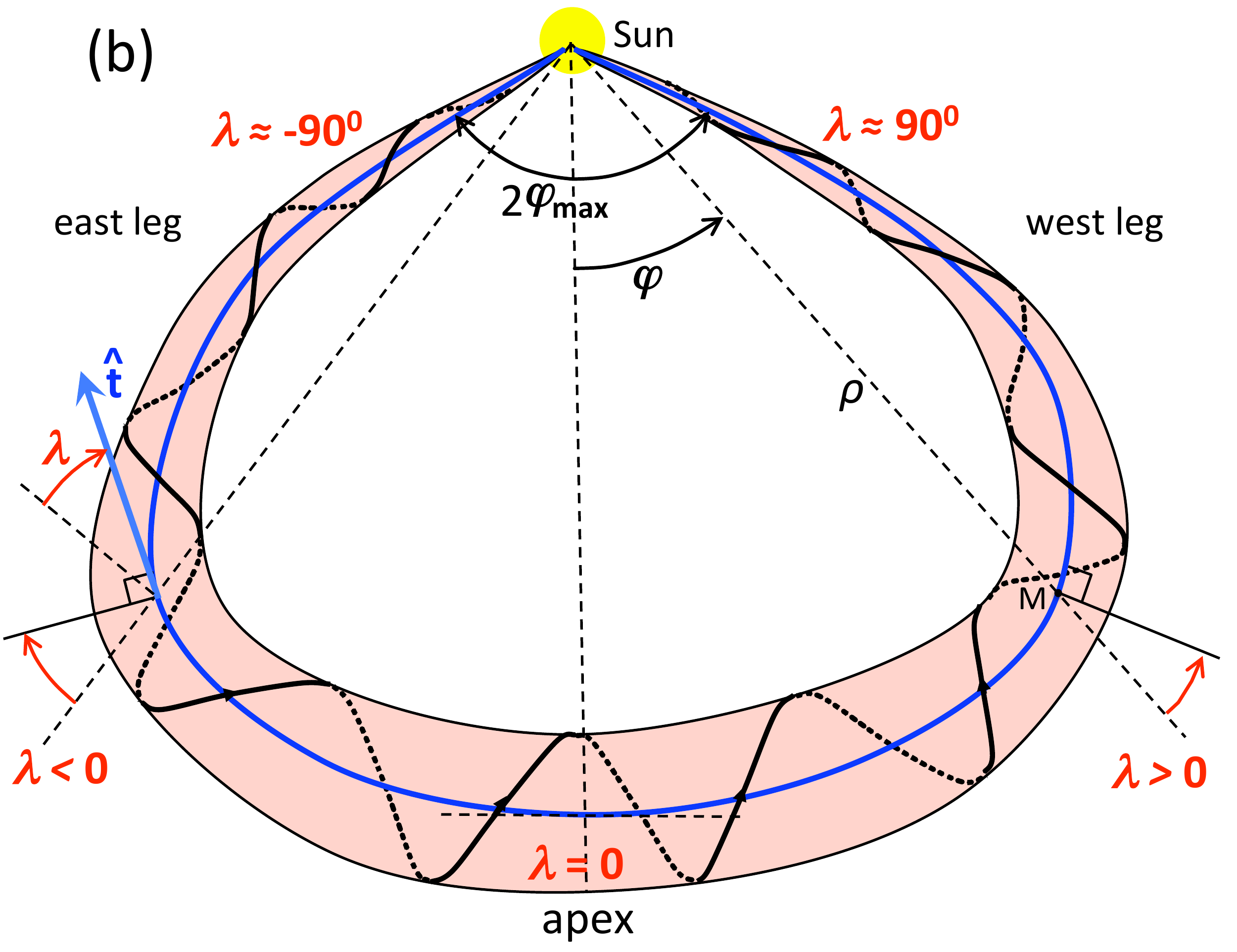} 
 }
\caption{
  (a) Graphic defining the inclination angle $\iA$ and the location angle $\lA$ of the local FR axis direction in the GSE system of coordinates.  In this diagram, $\iA>0$ and $\lA<0$ (corresponding to the east side of the FR shown on the right panel). $\uvec{t}$ is the unit vector tangent to the FR local axis.
  (b) Diagram showing the large scale meaning of $\lA$ for a FR launched from the Sun (and here still attached to the Sun to simplify the drawing).  The graphic is drawn in the plane of the FR axis, with an inclination $\iA$ on the ecliptic (left panel). $\varphi$ is the polar angle of cylindrical coordinates centred on the Sun and $\rho$ is the distance of the {point M on the FR axis} to the Sun.
}
 \label{fig:schema}
\end{figure}  

For almost North-South oriented FRs, \ie\ $|\tA| \approx 90\degree$, a small modification of the axis direction implies a large change of $\pA$. The angle $\pA$ is therefore not a reliable parameter for a fraction of the FRs, as a correlation study between FR parameters in function of $\pA$ would be misleading: the dispersion of $\pA$ would increase as $|\tA|$ is closer to $90\degree$ \citep[\eg,][]{Gulisano07}. Furthermore, North-South oriented FRs are present in the data sets, making the spherical coordinates $(\pA,\tA)$ not well adapted for the study of FR orientations.
Then, following \citet{Janvier13}, we set the polar axis of a new spherical coordinate system along $-\uvec{x}_{\rm GSE}$ (the radial direction away from the Sun). 
Since all the observed FR axes are far from this direction, the singularity of the spherical coordinates around the new polar axis is no longer a problem.  
With this new coordinate system, we define two angles: the inclination on the ecliptic $\iA$ and the location $\lA$ angles (\fig{schema}).
For a FR with its axis in a plane, the angle $\iA$ is the inclination of the FR plane with respect to the ecliptic (\fig{schema}a). For a FR with a simple shape, as shown in \fig{schema}b and in earlier publications \citep[\eg ][]{Burlaga90,Bothmer98}, the angle $\lA$ evolves monotonously along the FR, 
defining the location where the spacecraft intercepts the FR if the axis shape is known (\ie, parametrizing the global shape of the FR axis in function of $\lA$ links the value of $\lA$ measured from observations with the spacecraft crossing position along the axis).   

The available data sets only provide limited statistics. Then, as the FR shape is not expected to depend on the sign of the axial field, we set all axis directions, represented by the unit vector $\uvec{t}$, to point toward the east side by changing $\uvec{t}$ to $-\uvec{t}$ when needed. Then, $\pA$ is in the interval $[0\degree ,180\degree ]$. We checked the validity of this choice for both data sets within the limits of the statistics (\ie , the cases $\pA >0$ and $\pA <0$ have comparable properties).
To make the plots in function of $\lA$ easier to interpret, we change the sign convention of $\lA$ from \citet{Janvier13} so that the eastern (western) leg, corresponds to $\lA<0$ ($\lA>0$, \fig{schema}).  With this convention, plots in function of $\lA$ have the eastern data on the left side as one observes the Sun from Earth 
(\eg, \fig{shape_axis}).
Then, the relations between ($\iA ,\lA$) and ($\pA ,\tA$) are
\BA
\tan \iA &=& \tan \tA ~/~ \sin \pA  \label{eq:iA} \,,  \\
\sin \lA &=&-\cos \pA ~\cos \tA    \label{eq:lA} \,.
\EA 
Since $\sin \pA \ge 0$, the inclination angle $\iA$ has the same sign as the latitude $\tA$ (even more, for $\pA$ close to $90\degree$, so near the FR apex, $\iA \approx \tA$).  Both $\iA$ and $\lA$ are defined 
in the interval $[-90\degree , 90\degree ]$.

\section{Correlations Between Flux Rope Parameters} 
      \label{sect:Correlations} 

In this section, we investigate all the correlations between the FR parameters so as to derive some physical properties for the observed FRs.  Since the data sets provide a large number of FR parameters, we present in the following only the most meaningful correlations, and the cases that are not mentioned correspond to cases that are simply not significantly correlated.  We characterize the correlations by two coefficients: the Pearson ($\cP$) and the Spearman rank ($\cS$) correlation coefficients.  We also report in the figures the fit of the data with a straight blue line.

\begin{figure}  
 \centerline{ \includegraphics[width=0.5\textwidth]{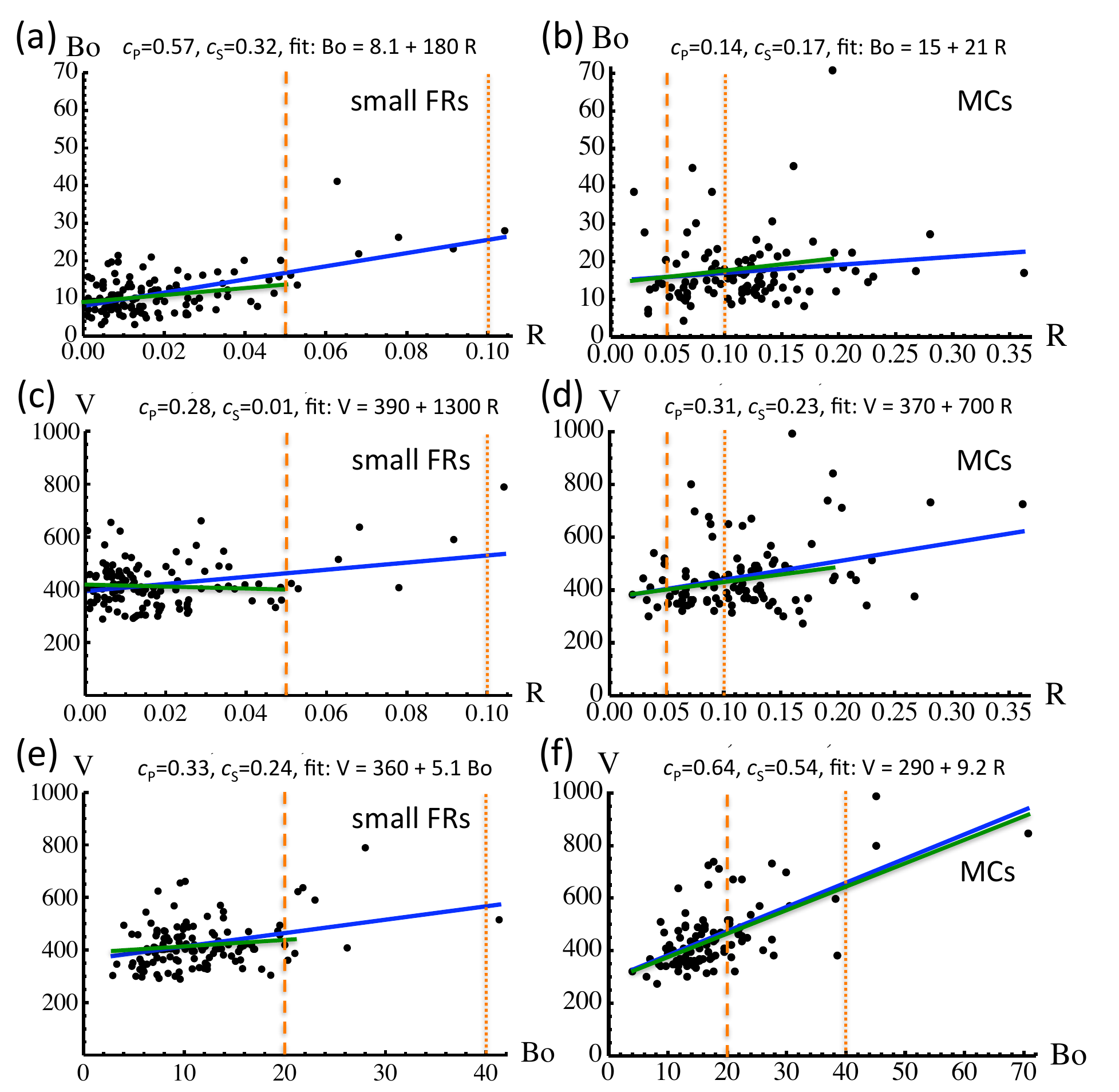} }
\caption{
  (a,b) Correlations between the axial magnetic field strength $\Bo$ (in nT) and the radius $\Ro$ (in AU). 
  (c,d) Correlations between the FR mean velocity $V$ (in \kms) and $\Ro$. 
  (e,f) Correlations between $V$ and $\Bo$.   The left column corresponds to the small FRs, and the right to MCs. The straight blue lines are linear fits to the data points showing the global tendency.  For small FRs (resp. MCs) the linear fits limited to $\Ro<0.05$ (resp. $\Ro<0.2$) AU are added in green color. The vertical dashed and dotted orange lines are guides to compare small FRs to MCs as the $\Ro$ and $\Bo$ ranges are different on the horizontal axis.}
 \label{fig:V(R)_Bo(R)}
\end{figure}  

\subsection{Correlations with Flux Rope Physical Parameters} 
      \label{sect:Correlations-General} 

\fig{V(R)_Bo(R)}a shows that for small FRs, there is a strong correlation between the axial field strength $\Bo$ and the FR radius $\Ro$. This correlation is also present in MCs but is much weaker (\fig{V(R)_Bo(R)}b).  There are few small FRs with $\Ro >0.05$ AU (only seven cases). Without these cases, both correlation coefficients decrease (to $\approx 0.24$), being closer to the values found for MCs.
  Similar results are found with the mean field strength $\aB$ along the spacecraft trajectory, showing that the correlations are not due to a bias of the FR fit to the data (assuming a Lundquist model could introduce a bias in the derived model parameters, \eg, as was shown by \citet{Demoulin13} for the impact parameter).
Finally, the small FRs typically have smaller axial field with a median $\Bo \approx 10$~nT, compared to the MCs with a median $\Bo \approx 16$~nT.  Both $\Bo$ distributions are Gaussian-like and they largely overlap as their standard deviations are $\approx 6$ and $7$~nT, respectively.
   
   Another correlation is found between the mean velocity $V$ and the FR radius $\Ro$ (\figsss{V(R)_Bo(R)}c,d), with a Pearson correlation coefficient similar for both populations of FRs ($\approx 0.3$). However, the Spearman coefficient indicates no correlation for small FRs. Furthermore, for the set of small FRs with radius $\Ro \leq 0.05$ AU, both Pearson and Spearman coefficients vanish. Also, the least square fits in blue and green show strong differences with and without small FRs with radius $\Ro > 0.05$ AU. This indicates that the correlation $V(\Ro)$ is weak for small FRs, as it is only present when the seven cases with $\Ro >0.05$ AU are included.  
Finally, small FRs and MCs travel away from the Sun with a speed typical of the slow SW ($V=420\pm 90$ and $450\pm 120$ \kms, respectively).     
     
  The mean velocity $V$ is also positively correlated with the axial field strength $\Bo$ (\figsss{V(R)_Bo(R)}e,f) with a correlation significantly stronger for MCs than for small FRs. 
Again, both the correlation parameters decrease (from mean values $\approx 0.29$ to $\approx 0.16$) when the condition $\Ro \leq 0.05$ AU is added for small FRs.   

{Similarly to small FRs, MCs have fewer cases with large $\Ro$ and $\Bo$ values (\figsss{V(R)_Bo(R)}b,d,f). Do these large events also affect significantly the correlations?} With the selection criterion $\Ro \leq 0.2$ AU, eight cases are removed, representing a {comparable fraction ($8/107 \approx 0.07$) of larger events} as for the small FRs ($7/121 \approx 0.06$). Similarly to small FRs, the most affected correlation is $V(\Ro)$. However, the effect is smaller since both correlation parameters only decrease from mean values $\approx 0.27$ to $\approx 0.21$. The effect of the criterion for the two other correlations of MC parameters is negligible (the change in the correlation coefficients is less than 0.02).  The correlation results are therefore more robust for MCs than for small FRs. Moreover, the selection on $R$ for MCs does not change significantly the mean tendency given by the least square fits (blue and green lines in \fig{V(R)_Bo(R)}).

 {The deviation between the observed magnetic field and the fitted FR model is characterized by $\sqrt{\chi}$.  For both small FRs and MCs, $\chi$ is below $0.3$ for all cases.  Its mean value is $0.18$ and $0.14$ for small FRs and MCs, respectively.  We investigate if the quality of the fit, so $\chi$, is function of the FR parameters to test if some types of FRs are less well fitted by the Lundquist's model.
 We find no significant correlation between the parameters derived from the fitted FR model and $\chi$ (the absolute values of the correlation coefficients are well below 0.2), except for the impact parameter where the correlation is $\approx -0.4$ and $\approx -0.14$ for small FRs and MCs, respectively. 
These negative values imply that the fit is better when detection is made closer to the FR boundary,
a result which was also found when the Lundquist's fit was tested with FR models \citep{Demoulin13}.   Finally, the absence of a significant correlation between the other FR parameters and $\chi$ implies that the derived FR properties are independent of the fit quality (with the restriction $\chi < 0.3$).} 

\subsection{Interpretation of the Correlations} 
      \label{sect:Correlations-Interpretation} 
   At this stage, some interpretations can already be given to explain the different correlation results. MCs are interplanetary consequences of CMEs ejected from the solar corona due to the destabilization of magnetic configurations (\eg, see the review of \citet{aulanier14}).  Events with a strong magnetic field strength have a large outward Lorentz force, so they are accelerated to large velocities. Then, a positive correlation is expected between field strength and mean velocity in the corona.  
   
   However, the CME properties are partly erased at a later stage of their propagation due to their interaction with the SW. For example, CMEs propagating faster than the ambient SW are deccelerated, while slower ones are accelerated.  Moreover, the FR radius evolves mostly as a consequence of the pressure balance between the FR and the surrounding medium (\sect{Expansion-Size}). Still, the correlations found at 1~AU (\fig{V(R)_Bo(R)}) indicate that the coronal conditions do not completely disappear during their propagation.
  
   The similar positive correlations found for small FRs indicate comparable physical processes for the two populations of FRs. Supposing that small FRs are also launched from the Sun, the difference in the correlation strength may be due to different destabilization mechanisms for small FRs launched from the corona, although transport effects in the SW can also have a key role in generating similar characteristics for both FR types.  {Indeed, \insitu\ observations closer to the Sun would be needed to evaluate the transport effects.}

\begin{figure}  
 \centerline{
   \includegraphics[width=0.5\textwidth]{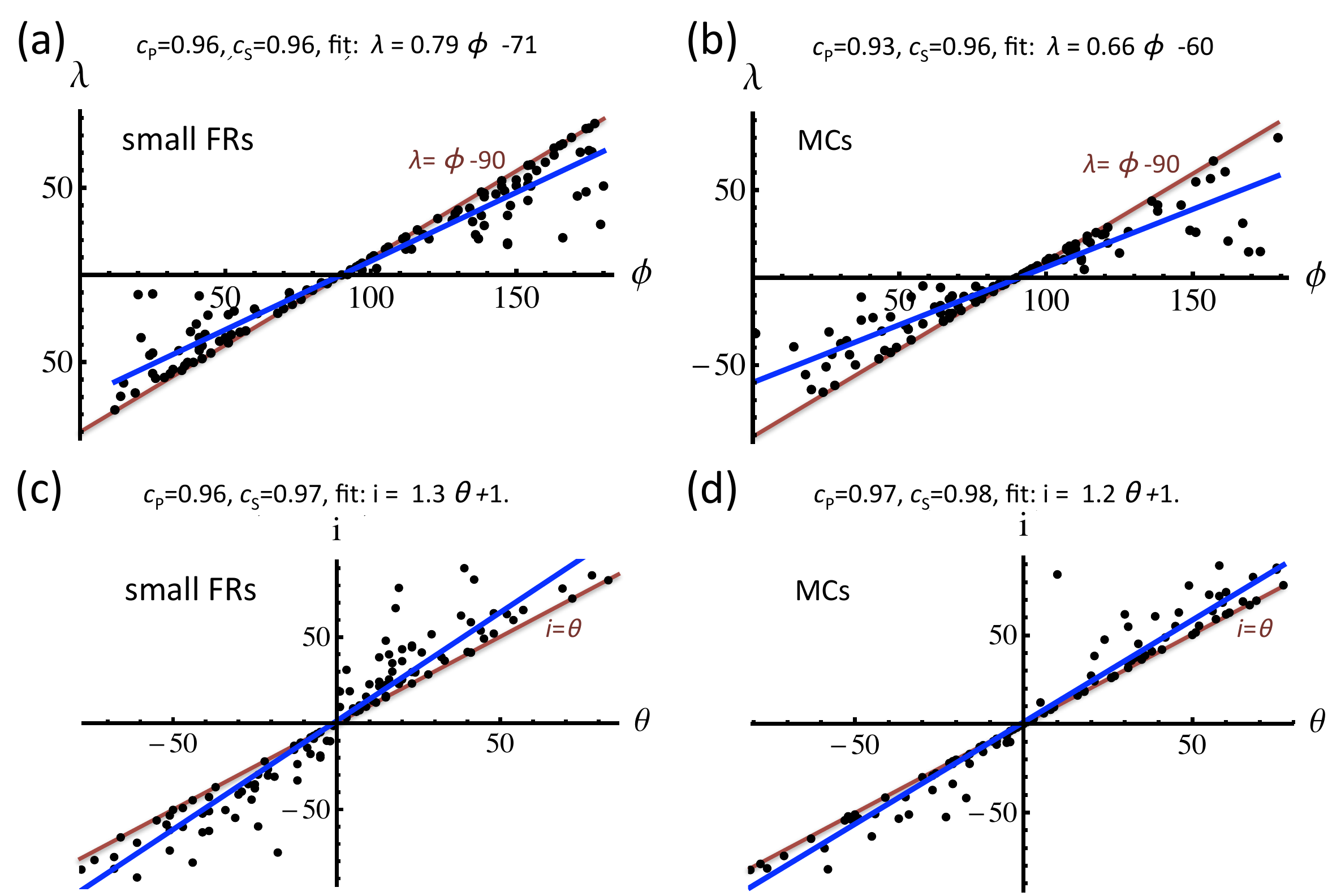}
 }
\caption{Correlations between the axis angles for small FRs (left column) and MCs (right column),
  (a,b) between the position angle $\lA$ and the longitude $\pA$,
  (c,d) between the inclination angle $\iA$ and the latitude $\tA$.
  $\lA$ and $\iA$ are defined in \fig{schema}. 
  The straight blue lines are linear fits to the data points, showing the global tendency and the straight brown lines are simple limit cases as indicated. 
}
 \label{fig:lA_iA}
\end{figure}  

\subsection{Correlations between Axis Orientation Angles} 
      \label{sect:Correlations-Axis} 

 {The} inclination $\iA$ and the location $\lA$ angles are related to the longitude $\pA$ and the latitude $\tA$ with \eqs{iA}{lA}.  
However, $\iA$ has a low correlation with $\pA$ (correlation coefficients $\approx 0.07$ for small FRs and $\leq 0.03$ for MCs) and $\lA$ has a low correlation with $\tA$ 
(correlation coefficients $\approx 0.05$ for small FRs and $\approx 0.02$ for MCs).   On the contrary, $\lA$ is very well correlated with $\pA$ and $\iA$ with $\tA$ for small FRs and for MCs with very similar correlation coefficients ($\geq 0.93$, \fig{lA_iA}).

These results are explained as follows.
The FRs with low $\tA$ values are located close to the straight brown line $\lA \approx \pA -90$ (\figsss{lA_iA}a,b).  The FRs with larger $\tA$ values have lower $|\lA|$ because $\cos \tA < 1$ in \eq{lA}.  This implies that the slope of the linear fit (blue line) is lower than 1 ($=0.79$ for small FRs, $=0.66$ for MCs).  
Similarly, the FRs with $\pA$ around $90 \degree$ ($y_{\rm GSE}$ direction) are displayed along the brown line $\iA \approx \tA$ in \figsss{lA_iA}c,d.  The FRs away from the $y_{\rm GSE}$ direction have lower $\sin \pA$, then $|\iA|$ is larger than $|\tA|$, as deduced from \eq{iA}.  This implies that the slope for the linear fit of $\iA (\tA)$ is larger than 1 ($=1.3$ for small FRs, $=1.2$ for MCs). These correlations are worth noticing since they give much clearer relations between the newly introduced angles $\iA$ and $\lA$, and $\pA$ and $\tA$ than \eqs{iA}{lA}.

We find no correlation between any of the FR parameters and the inclination angle $\iA$ for both small FRs and MCs ($|\cP|$ and $|\cS|$ are both below $0.15$).
This implies that both sets of FRs have properties independent of their axis orientation around the Sun-spacecraft direction.  {This is an important step to check, as we did not know \textit{a priori} whether the inclination of the flux ropes on the ecliptic could affect their properties (for example, the FR interaction with the SW magnetic field could a priori introduce a dependence on the inclination angle $\iA$). This also shows that small flux ropes are not necessarily located in the ecliptic plane, and can therefore be found away from the HCS, as suggested by the varied range of inclination angle $i$ found here.}

  {Finally,} we group both North and South hemispheres by analyzing correlations with the parameter $|\iA|$. There is only a weak correlation (coefficients $\approx 0.2$) of $\Ro$ with $|\iA|$ for small FRs but no other significant correlation is found.  So, similarly to MCs \citep{Janvier13}, 
we conclude that the small FR properties are independent of the FR inclination on the ecliptic.

\begin{figure}  
 \centerline{ \includegraphics[width=0.5\textwidth]{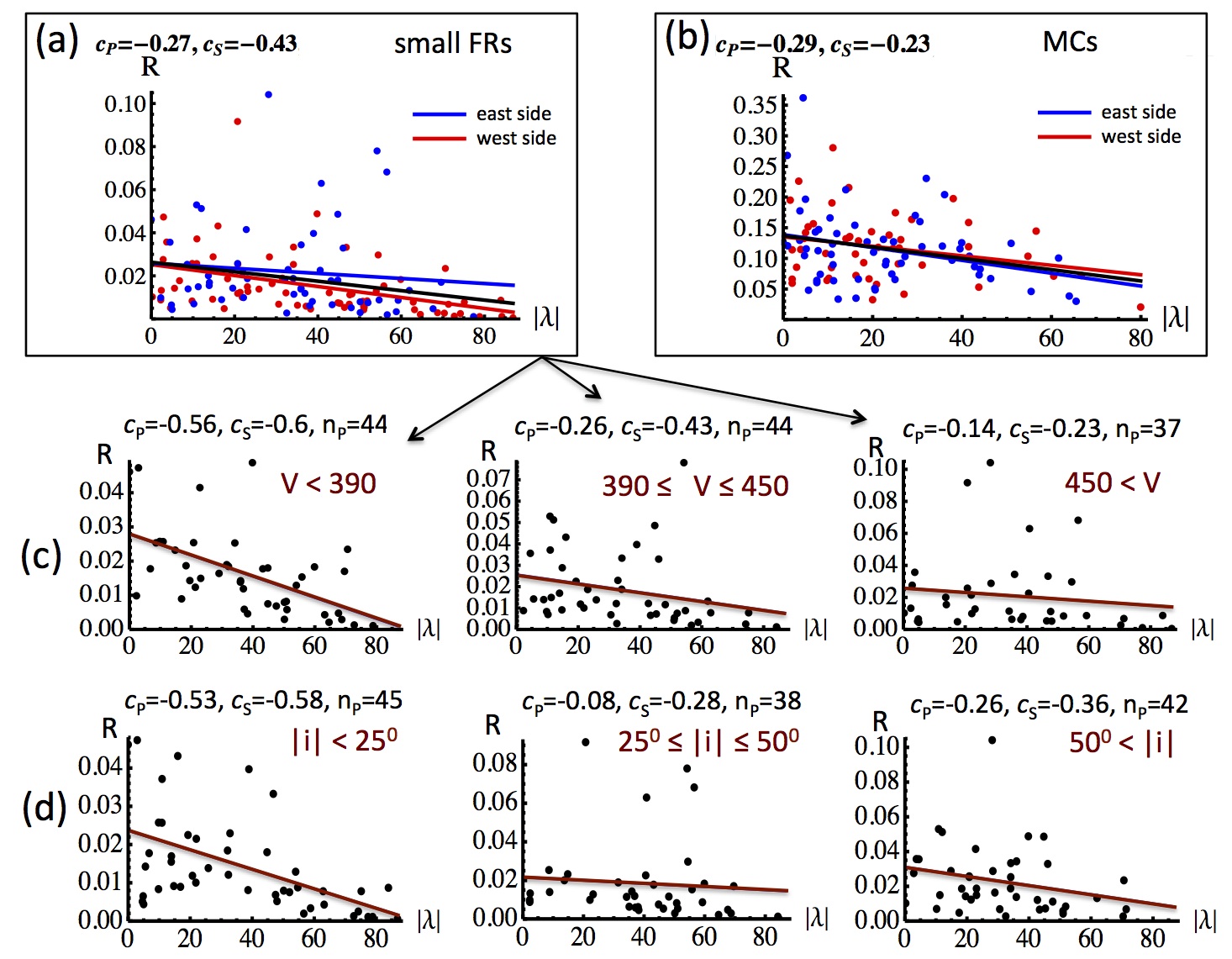} }
\caption{
(a) Correlations of FR radius $\Ro$ (in AU), with the {unsigned} location angle $|\lA |$ (in degree), for the small FRs of Feng et al (2008).
(b) Correlations of $\Ro$ with $|\lA |$ for the MCs of Lepping and Wu (2010).
$\lA >0$ and $\lA <0$ are respectively shown in red and blue, while the straight lines are linear fits to the data points showing the global tendencies.
The results with the total set are shown in black (linear fit and top labels).
(c) The correlations of $\Ro(\lA )$ for small FRs are separated in three groups of velocity $V$ (in \kms ) with similar number of cases. 
(d) Same as (c) but with a separation in three groups of inclination $|\iA |$. 
For panels (c,d) the straight brown lines are linear fits to the data points.}

 \label{fig:R_apex_legs}
\end{figure}  

\subsection{Mean Variations along the Flux Rope Axis} 
      \label{sect:alongFR} 

The parameter $\lA$ provides an information on the location along the FR axis if we suppose that the FRs crossed by the \textit{Wind} spacecraft have a typical shape, curved towards the Sun, as the one shown in \fig{schema}.
In such a case, correlation plots of FR parameters in function of $\lA$ indicate the mean variations along the FR axis. In particular, these plots allow us to check whether there is an asymmetry between the east and the west legs. Similarly, correlations with $|\lA|$ allow us to compare the mean properties of the apex region with the leg {regions.} 

  All the absolute values of the correlation coefficients with signed $\lA$ are $\le 0.1$ for MCs, so there is no significant difference between the MC legs  (as shown in \fig{R_apex_legs}b by comparing the blue and red straight lines). For small FRs, the highest correlation found is a negative one between $\Ro$ and signed $\lA$, so that the western side is thinner than the eastern one. This result is shown in \fig{R_apex_legs}a with $\Ro(|\lA|)$ with the difference between the blue and red straight lines.  
Still, both values for the Spearman and Pearson coefficients of $\Ro(\lA)$ are small ($\approx -0.24$). Furthermore, we also notice that removing from the sample the cases with $\Ro > 0.05$ AU (seven small FRs) leads to a decrease in the correlation coefficients (to $\approx -0.18$). Therefore, the correlation is enhanced by these few cases and we conclude that within the limited sample of small FRs studied, there is no significant difference between the two legs of the small FRs, similarly to MCs.

The correlation plots with $|\lA|$ allow us to compare the averaged properties at the apex with those at the FR legs.  Both for small FRs and MCs, there is no other global tendency than a bigger radius at the apex compared to the legs (\figsss{R_apex_legs}a,b).   
Contrary to the correlation discussed above for $\Ro (\lA)$, this {$\Ro(|\lA|)$} correlation is strong. For example, removing from the sample the cases with $\Ro \geq 0.05$ AU in \fig{R_apex_legs}a, both correlation coefficients become even more negative (with values around $\approx -0.44$). 

The above result is not linked to the popular cartoon of a FR getting thinner toward the legs (\eg\ \fig{schema}b). Indeed, this cartoon is done at {one} given time, {so with a radius varying with the variable solar distance} along the FR. {This means that at one given time, the apex, which would have propagated farther away from the Sun than the legs, would appear more extended than the legs. However, here,} all FRs are observed at 1~AU, {independently of where they are crossed along their axis. Then, flux ropes detected at their legs would have propagated for a longer time than flux ropes that are detected at their apex, and both legs and apex, at 1AU, would be expected to have a similar radius due to a balance of pressure with the surrounding SW (see \sect{Expansion-Size})}. We found that $\Bo$ and $\aB$ have similar values at the apex and in the legs (not shown), as indeed expected with a balance of pressure between the FR and the surrounding medium at the same solar distance. 
It implies that a thinner radius in the legs than at the apex is not due to compression.

   Next, we explore whether the above $R(|\lA|)$ correlation is a function of the other FR properties by dividing the data in subgroups of FRs with more homogeneous properties. To do so, we divide the small FRs and MCs samples in groups of $\approx 40$ FRs with similar number of data points $\rm n_p$, to have comparable statistical fluctuations. We limit the division to three groups so as to avoid statistical fluctuations ($1/\sqrt{\rm n_p}\approx 0.16$ for $\rm n_p \approx 40$). We apply this grouping successively to the FRs ordered by any of the other FR parameters.
The tendency of an apex broader than the legs increases for low $V$ and low $|\iA|$ for small FRs (\fig{R_apex_legs}c,d) and for larger $\aB$ (while the FRs remain broader at the apex for any of the $\Bo$ values).  For MCs, the strongest effect is for $\Bo$ and $\aB$: the weaker-field MCs show no correlation between $R$ and $|\lA|$, while the strong-field MCs have both correlation coefficients of the order $\approx -0.4$.  For MCs, there is only a weak tendency for $|\iA|$ with correlation coefficients for $R(|\lA |)$ of the order of $\approx -0.34$ for low $|\iA|$ and $\approx -0.25$ for large $|\iA|$ (with the same three groups {of $|\iA|$} than in \fig{R_apex_legs}d). 

In summary, both small FRs and MCs have thinner legs than the apex in average, and this effect is stronger for lower $V$, lower $|\iA|$ and larger $\aB$.  However, the dominant effect is due to $V$ and $|\iA|$ for small FRs while it is due to $\aB$ and $\Bo$ for MCs.

\subsection{Interpretation of a Variable Radius Along the FR Axis} 
      \label{sect:Variable-Radius} 

In this section, we investigate the origin of the radius difference between the apex and the leg region for FRs crossed at 1~AU.

A first possibility is an increasing bias in the Lundquist's fit as the spacecraft crossing is further away from the apex.  When the spacecraft trajectory is crossing the FR less perpendicularly to its axis (larger $|\lA|$), the spacecraft trajectory explores a longer part along the FR, so the magnetic field measured is more affected by the bending of the MC axis.  It implies that the hypothesis of a local straight FR, used in the Lundquist model, is less valid as $|\lA|$ increases \citep{Owens12}.  This could bias the radius estimation from the Lundquist's fit.  However, such a bias should be independent of $V$ and $|\iA|$, contrary to what is found above. Then, we conclude that the fitting model bias cannot fully explain the results presented above.
  
A second possibility is a physical origin during the interplanetary transport, that could explain the puzzling asymmetry in MCs between the apex and the legs for stronger magnetic field, and the increase of the asymmetry for lower velocities for small FRs. 
{Previous works have empirically shown that magnetic reconnection at the front boundary of a MC \citep[\eg,][]{Dasso06,Dasso07,Ruffenach12,Lavraud14} or a small flux rope \citep{Tian10} can erode parts of their magnetic structure}. 
Then, a possible explanation for the different FR size at the legs/apex is an efficient FR erosion via magnetic reconnection.
The FR as an entity is launched at a given time, and its nose reaches 1 AU sooner than the legs. Then, the legs end up spending more time travelling in the solar wind than the nose, \ie, they spend more time in conditions favouring the erosion process than the nose.
This scenario could explain the $|\lA|$ correlation with $V$, since a lower velocity implies more time for the erosion to take place. This effect however competes with a less efficient reconnection process when the relative velocity with the solar wind speed is lower, so that the correlation with $V$ still remains unclear.

Finally, a third possibility is of coronal origin, as follows.  During the quasi-static build-up of the coronal FR, and then even more during its eruption, part of the overlaying magnetic arcade is forced to reconnect below the FR (\eg , \citet{Aulanier10}). Then, the overlaying arcade field is progressively included as part of the FR.  
The magnetic shear of the reconnecting arcades is typically decreasing with time (\eg , \citet{Aulanier12} and references therein). At larger altitudes, these arcades are therefore more orthogonal to the FR axis. These weakly sheared arcades are still forced to reconnect in the trailing reconnecting region below the FR, further building up its envelope around the apex. Away from the FR apex, arcade field lines would simply be pushed away by the expanding flux rope (as shown in Figure 2 of \citet{Sturrock2001}), therefore not creating more flux in the leg regions.

In conclusion, having thinner legs than the nose region in the interplanetary medium is probably a consequence of both coronal and interplanetary processes.
The coronal processes are expected to affect both MCs and small FRs {(if formed in the corona)}. However, since for small FRs the difference between the apex and leg regions depends on $V$ and $|\iA|$, the interplanetary processes seem to strongly affect small FRs, while the coronal ones are dominant for MCs. This conclusion needs to be further confirmed by future studies.

\section{Statistical Properties of the Axis Orientation} 
      \label{sect:Orientation} 

  In this section, we analyse the distributions of the $\iA$ and $\lA$ angles.  If the small FRs are formed in the interplanetary medium by local reconnection within a current sheet, as advocated by \citet{Cartwright08},  the distributions would provide some information on the properties of the original magnetic field involved (which defines, after reconnection in the current sheet, the axial and azimuthal magnetic components of the FR).  Instead, if the small FRs are launched from the Sun, the distributions of $\iA$ and $\lA$ can provide a statistical derivation of the mean FR axis shape as achieved previously for the MCs \citep{Janvier13}, and as further developed in \sect{Orientation-Shape}.  
      
\begin{figure}  
 \centerline{ \includegraphics[width=0.5\textwidth]{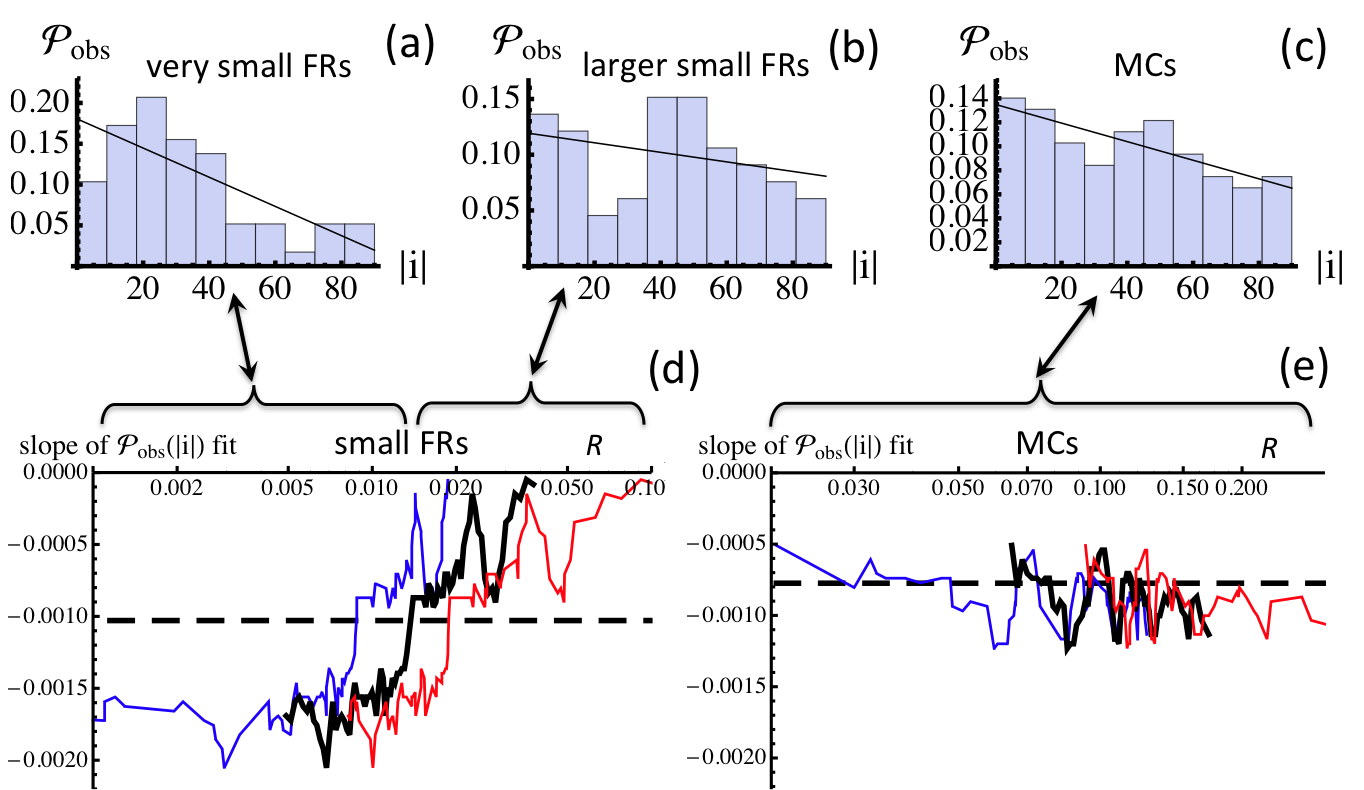} }
\caption{Dependence of the slope of the probability distribution $\pobsAi$ with the FR radius $\Ro$ for small FRs and MCs.
(a-c) Three histograms of $\pobsAi$ for selected ranges of $\Ro$.
The slope of the linear fit (straight black line) is strongly dependent on the $\Ro$ range as indicated by the horizontal braces.  
(d,e) The evolution of this slope is shown, with FRs grouped in subsets of 40 FRs, shifting progressively $\Ro$ to larger values (see \sect{Orientation-Inclination}).
The three curves represent the slope of the fit, with the black line corresponding to the mean value of $\Ro$ for each subset, and the blue (resp. red) line corresponding to the minimum (resp. maximum) of $\Ro$ for each subset. The horizontal dashed line is the slope of the probability distribution $\pobsAi$ for all the FRs in the panel.
}
 \label{fig:slope_Pi}
\end{figure}  

\subsection{Distribution of the Inclination Angle $\iA$} 
      \label{sect:Orientation-Inclination} 

  The distribution of $\iA$ shows a tendency of finding less small FRs when they are oriented in the North/South direction ($|\iA|$ close to $90 \degree$). This effect is stronger for smaller radius (\figsss{slope_Pi}a,b). Therefore, the axis of very small FRs is more frequently parallel to the ecliptic compared to MCs, which have more uniformly distributed $|\iA|$ values (\fig{slope_Pi}c). {MCs have a weak tendency to be more frequently parallel to the ecliptic at 1~AU, in contrast with previous results obtained with Helios spacecraft between 0.3 and 1~AU \citep[Figure 9 of ][]{Bothmer98}.}

  In order to further study this radius dependence, we first order the FRs by growing values of $\Ro$. We then make histograms, as shown in \figsss{slope_Pi}a-c, with first a subset of the $n_{\rm s}$ smallest FRs. Then, we iterate the analysis by removing the smallest FR and by adding the next larger FR. As such, we change from one subset to another subset of $n_{\rm s}$ FRs and progressively scan all the FR set.   The fit of each histogram provides the slope as a function of the mean value of $\Ro$ in each FR subset (black line in \figsss{slope_Pi}d,e).   We also show the slope versus the minimum (maximum) value of $\Ro$ for each subset with blue (red) lines, respectively, to indicate the range of $\Ro$ involved in the slope determination.   We show the results for $n_{\rm s}=40$ as a compromise between decreasing the statistical fluctuations and having enough resolution in $\Ro$.  

   \fig{slope_Pi}d shows that the slope of $\pobsAi$ is a monotonously increasing function of $\Ro$:
the very small FRs are dominantly closer to the ecliptic (large negative slope as in \fig{slope_Pi}a) while the larger small FRs are nearly equi-distributed in $|\iA|$ (as in \fig{slope_Pi}b). 
Repeating the above procedure with subsets of MCs, we found no effect of $\Ro$ on the slope of $\pobsAi$ (\fig{slope_Pi}e), so that MCs are also nearly equi-distributed in $|\iA|$ (as \fig{slope_Pi}c).

\subsection{Interpretation of the $\iA$ Angle Distribution} 
      \label{sect:Interpretation-Inclination} 
      
{Following the previous analysis, we analyse }why the $|\iA|$ distribution of small FRs is changing with the FR radius (\fig{slope_Pi}d).  Smaller FRs are expected to be more affected by the surrounding SW than larger ones.  Since the small FRs are typically found nearby the HCS \citep{Cartwright10}, their orientation could be affected by the magnetic field of surrounding magnetic sectors.

The HCS typical scale is $\le 10^{-4}$~AU \citep{Winterhalter94}, which is too small compared to the small FR radius (\fig{slope_Pi}) to have a significant effect.  However, the HCS is embedded in the heliospheric plasma sheet (HPS), which is characterized by an enhanced plasma $\beta$ and density with a drop in the magnetic field strength to ensure an approximate total pressure balance.  The HPS thickness at 1~AU was found to be below 0.01 AU by 
\citet{Winterhalter94}, while larger values, in the interval $[0.01,0.03]$~AU, were measured by \citet{Bavassano97} and \citet{Zhou05}.  These larger thickness values were confirmed by \citet{Foullon09} and \citet{Simunac12} with the crossing of the same HPS by three spacecraft during five solar rotations. They found that the HPS thickness does not significantly evolve on the time scale of a day, but does evolve at the scale of a solar rotation, with a thickness evolving between $0.02$ and $0.06$~AU.  

Altogether, the above range of HPS half thickness ($[0.005,0.03]$~AU) corresponds to the interval of small FR radius where the slope of $\pobsAi$ is significantly changing (\fig{slope_Pi}d). As such, we propose as an explanation that the smaller FRs are channelled within the HPS with the magnetic pressure of the surrounding SW forcing a global rotation of smaller FRs. These FRs then become nearly parallel to the HPS that is, in average, around the ecliptic plane. {The smaller FRs are also easier to rotate as their magnetic field and flux are weaker.
A} significant FR axis rotation (larger than $40\degree$), was also found for three CMEs as they traveled from 30 $R_{\odot}$ to 1~AU \citep{Isavnin14}. With 14 slow CMEs, observed at 1~AU as MCs, they also found a tendency for FR axis to be aligned with the HCS computed from an MHD simulation having magnetic synoptic maps as solar boundary conditions. The proposed mechanism for such a rotation is the interaction with the magnetic field and the streams of the solar wind. Such an interaction also implies a significant deflection from a radial trajectory \citep{Lugaz10,Kilpua09b,Isavnin13}. 

   { Our statistical results} indicate that the interaction has a stronger effect on small FRs than on MCs.   This can be explained as follows. We consider a small FR with a diameter smaller than the HPS thickness, and with an axis un-parallel to the HPS.  As the FR propagates away from the Sun, it should push the ambient magnetic field away or reconnect with it. The second case could be at the origin of less magnetic flux in the legs ({\sectss{alongFR}{Variable-Radius}}). The first case occurs if reconnection is not efficient enough. Then, a built-up of magnetic pressure occurs on the external sides of the FR, creating a torque that tends to rotate the FR axis parallel to the HPS. As the FR diameter increases above the HPS thickness, this process is only effective on a smaller fraction of the FR boundary. Then, larger FRs are simply not sensitive to the localised magnetic gradient present within the HPS.  Moreover, since the magnetic flux increases as the square of the diameter, larger FRs are more difficult to deform. We propose this mechanism as the origin of the transition of $\pobsAi$ with $\Ro$ shown in \fig{slope_Pi}. 

An alternative interpretation for the origin of the very small FRs {has been proposed as follows. Some authors have discussed a possible formation of small FRs} by reconnection at the HCS \citep{Cartwright08}{. Following their formation}, they would be parallel to the HPS. However, larger small FRs have a more spread distribution of $\iA$ (\fig{slope_Pi}b), comparable to MCs (\fig{slope_Pi}c) while all small FRs are thought to have the same origin (in particular, they have a common power-law distribution of sizes, \citet{Janvier14}).   These results are instead compatible with a solar origin of small FRs{, and it seems that an origin in the heliospheric current sheet would be unlikely.}
Small FRs are therefore also expected to be launched with a flat distribution of $|\iA|$, similarly to MCs, and only the smaller ones are rotated and channelled by the HPS. 
    
\begin{figure}  
 \centerline{ \includegraphics[width=0.5\textwidth]{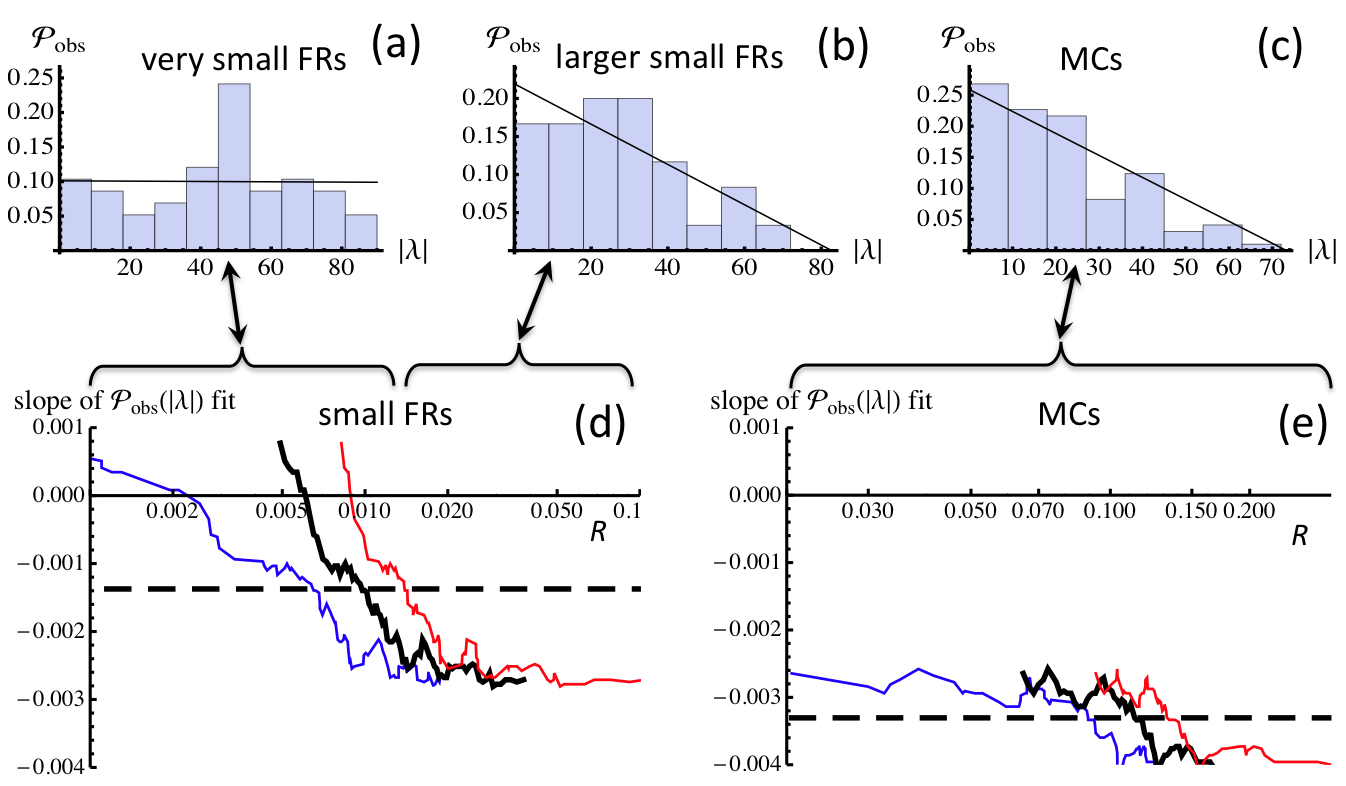} }
\caption{Dependence of the probability distribution $\pobsAl$ with the FR radius $\Ro$ for small FRs and MCs.
  (a-c) Three histograms of $\pobsAl$ for selected ranges of $\Ro$.
The slope of the linear fit (straight black line) is strongly dependent on the $\Ro$ range as indicated by the horizontal braces. 
  (d,e) The evolution of this slope is shown in the top panels where the FRs are grouped in subsets of 40 FRs, shifting progressively $\Ro$ to larger values (see \sect{Orientation-Location}).
The three curves represent the slope of the fit, with the black line corresponding to the mean value of $\Ro$ for each subset, and the blue (resp. red) line corresponding to the minimum (resp. maximum) of $\Ro$ for each subset. The horizontal dashed line is the slope of the probability distribution $\pobsAl$ for all the FRs in the panel.
} 
 \label{fig:slope_Plambda}
\end{figure}  

\subsection{Distribution of the Location Angle $\lA$} 
      \label{sect:Orientation-Location} 

   The observed distribution of $|\lA|$, {$\pobsAl$,} is a steeply decreasing function for MCs (\fig{slope_Plambda}c and \citet{Janvier13}).  The slope is slightly weaker for the set of small FRs with larger radius (\fig{slope_Plambda}b) and it almost vanishes for the very small FRs (\fig{slope_Plambda}a). 
In Section \ref{sect:Orientation-Inclination}, we have shown how to further study the radius dependence of $\pobsAi$ by analyzing the slope of histograms made with a number $n_{\rm s}\; (=40)$ of FRs ordered by growing value of $\Ro$. Applying this technic to the distribution of $|\lA|$, \fig{slope_Plambda}d shows that the slope of $\pobsAl$ is a monotonous decreasing function of $\Ro$  {for small FRs.}  A weaker tendency is also present for MCs (\fig{slope_Plambda}e and \citet{Janvier13}).  

Let us first consider the hypothesis that small FRs are formed in the HCS by reconnection between the magnetic field of the opposite sectors present on both sides of the HCS. In such a case, one would expect a privileged direction of  {the FR axis: orthogonal to the local Parker spiral,}
at about $45 \degree$ westward from the radial direction at 1~AU for the slow SW (so $\lA \approx 45 \degree$, see \fig{schema}).  In \fig{slope_Plambda}a, a peak is indeed present for $|\lA|\approx 50 \degree$, corresponding to $\lA \approx 50 \degree$ (\fig{shape_axis}a).
This, along with the previous finding of small FR axis having a tendency to be parallel to the ecliptic, points toward an \insitu\ formation of small FRs. However, there are only 12 cases in the $\lA \approx 50 \degree$ peak, and the mean of the histogram is 6 cases per bin. This implies that only 6 small FRs would be oriented as expected from a formation process involving reconnection at the HCS.
 
   Moreover, $\pobsAl$ for very small FRs has also a broad extension, nearly uniform in all other $|\lA|$ bins (\fig{slope_Plambda}a).  Furthermore, for larger small FRs (\fig{slope_Plambda}b), $\pobsAl$ is more similar to $\pobsAl$ of MCs (\fig{slope_Plambda}c) and there is indeed a continuous evolution of the slope of $\pobsAl$ from very small FRs to large MCs (\fig{slope_Plambda}d,e).        
These arguments favor the scenario of small FRs ejected from the corona, similarly to MCs, but not necessarily with the same mechanisms ({e.g., formation with jets and coronal streamers, as discussed by \citet{Janvier14}).}  
The change of the slope of $\pobsAl$ occurs for the same range of radius, around $0.01$~AU, as for $\pobsAi$ in \fig{slope_Pi}d.  Since $\pobsAl$ is directly related to the shape of FR axis (\sect{Orientation-Shape}), it implies that the {axis} shape of the smaller FRs is affected by the interaction with the HPS.  More precisely, the larger small FRs are not significantly affected by the HPS, similarly to MCs, while the smaller ones, with sizes of the order or smaller than the size of the HPS, {can have an axis} deformed by the interaction with the surrounding sector magnetic fields, and the HPS with high plasma $\beta$.
   
   A further argument against a possible formation of small FRs in the HCS is related to the polarity of the small FRs, defined as the temporal ordering of the North/South component ($B_{z}$ field component in GSE coordinates) during the FR crossing. More precisely, a fraction of the small FRs has $B_{z}$ directed to the north in the leading part and to the south in the rear part (they are called NS-type), while the other FR fraction is of the SN type \citep{Cartwright08}.  If FRs are formed in the HCS, one would expect that the dominant type changes when the global solar dipole changes sign (typically one year after the solar maximum), as it implies that the magnetic field on both sides of the HCS changes sign.  Then, the absence of a NS/SN dominance for small FRs well before/after the global magnetic field reversal of the Sun \citep[Figure 8 of][]{Cartwright08} goes against the formation of small FRs by reconnection at the HCS. 

\subsection{Interpretation of the $\lA$ Angle Distribution: Shapes of the Flux Rope Axis} 
      \label{sect:Orientation-Shape} 

The discussions above point toward a coherent understanding of a coronal formation of small FRs, similarly to MCs. With the hypothesis that they have a coherent FR shape kept up to 1~AU, as shown in \fig{schema}b (but not necessarily still attached to the Sun), the results described in \sect{Orientation-Location} on $\pobsl$ can be used to obtain a mean FR shape, as shown in \citet{Janvier13} for MCs, and extended to asymmetric cases below.

We suppose that the FR axis of any analysed FR is located in a plane inclined by an angle $\iA$ on the ecliptic plane (\fig{schema}). Then, the FR axis can be expressed in cylindrical coordinates ($\rho , \varphi$) within the FR plane as 
  \BE \label{eq:OM}
  \vec{SM} = \rho (\varphi) ~\ur  \, ,
  \EE
where $\vec{SM}$ is from the Sun (S) to the point $M$ of the axis and $\ur$ is the unit vector in the radial direction.
We further make two hypotheses partly justified by present heliospheric imager data of CMEs, \ie\ for MCs, while small FRs are typically not imaged (small blobs from helmet streamers have been imaged, such as in \citet{Rouillard10} with possible connections to MCs \citep{Rouillard10b}, but small FRs detection is only possible on the rare cases of a sufficiently enhanced plasma density). A first hypothesis is that the full angular extension, as seen from the Sun, is $2\;\phimax$ (\fig{schema}).  A second hypothesis is that $\lambda$ is a monotonous function of $\varphi$ with $\lambda$ growing from $-90\degree$ to $90\degree$, from the east leg to the west one.  This implies that $\rho$ is a decreasing function of $|\varphi|$ from the axis apex to any of the legs (see \eq{tanLambda} below) so that the axis has a shape as shown in \fig{schema}b. 

The probability of a spacecraft crossing a FR with an angle $\varphi$ in a range $\rmd \varphi$ is defined by $\pvphi (\varphi) \rmd \varphi$.   Similarly, the probability of a spacecraft crossing a FR with the location angle $\lA$ in a range $\rmd \lA$ is $\pobsl \rmd \lA$, with $\pobsl$ provided by the FR observations.  The conservation of the number of cases implies
  \BE \label{eq:conservation}
  \pvphi (\varphi ) ~\rmd \varphi = \pobsl ~\rmd \lA  \, .
  \EE
Since the CMEs are launched from a broad range of solar latitudes and longitudes over the time scale of the analysed MC set (almost a solar cycle), the MCs are expected to be crossed with a uniform distribution in $\varphi$. Therefore, $\pvphi (\varphi )$ is independent of $\varphi$. Supposing a similar solar origin for small FRs, we expect that the {above} hypotheses still hold for the small FRs.  Since we suppose that the separation angle of the MC legs is $2~\phimax$, the normalization of the probability to unity implies $\pvphi = 1/(2\phimax)$, and \eq{conservation} becomes:  
  \BE \label{eq:dvarphi}
  \rmd \varphi = 2 \phimax \pobsl ~\rmd \lA  \, .
  \EE

The integration of \eq{dvarphi} provides $\varphi$ as a function of $\lA$ as 
  \BE \label{eq:varphi}
  \varphi (\lA) = 2 \phimax \int_{0}^{\lA} \pobs (\lA') ~\rmd \lA'  \, .
  \EE
Next, we relate $\rho$ to $\lA$ by computing the unit tangent vector $\uvec{t}$ to the axis  at the point M {(\fig{schema}).} Then, we express $\lA$ as the angle between $\uvec{t}$ and the ortho-radial direction, $\up$, which writes as  
  \BE \label{eq:tanLambda}
  \frac{\rmd \ln \rho}{\rmd \varphi} = - \tan \lA  \, .
  \EE
Using \eq{dvarphi},  the integration of \eq{tanLambda} implies
  \BE \label{eq:rho}
  \ln \rho (\lA) = -2 \phimax  
            \int_{0}^{\lA} \tan (\lA') ~\pobs (\lA') ~\rmd \lA'
           +\ln \rho_{\rm max}  \, .
  \EE
\eqs{varphi}{rho} define a mean FR shape as a parametric curve ($\rho (\lA), \varphi (\lA)$) in cylindrical coordinates, from the observed probability distribution $\pobsl$ of FRs.  The axis shape depends on the two constants $\phimax $ and $\rho_{\rm max}$.  The second one is fixed to 1~AU as the small FRs and MCs are observed by the Wind spacecraft, while $\phimax $ cannot be defined by \insitu\ observations.

\begin{figure}  
\centerline{ \includegraphics[width=0.5\textwidth]{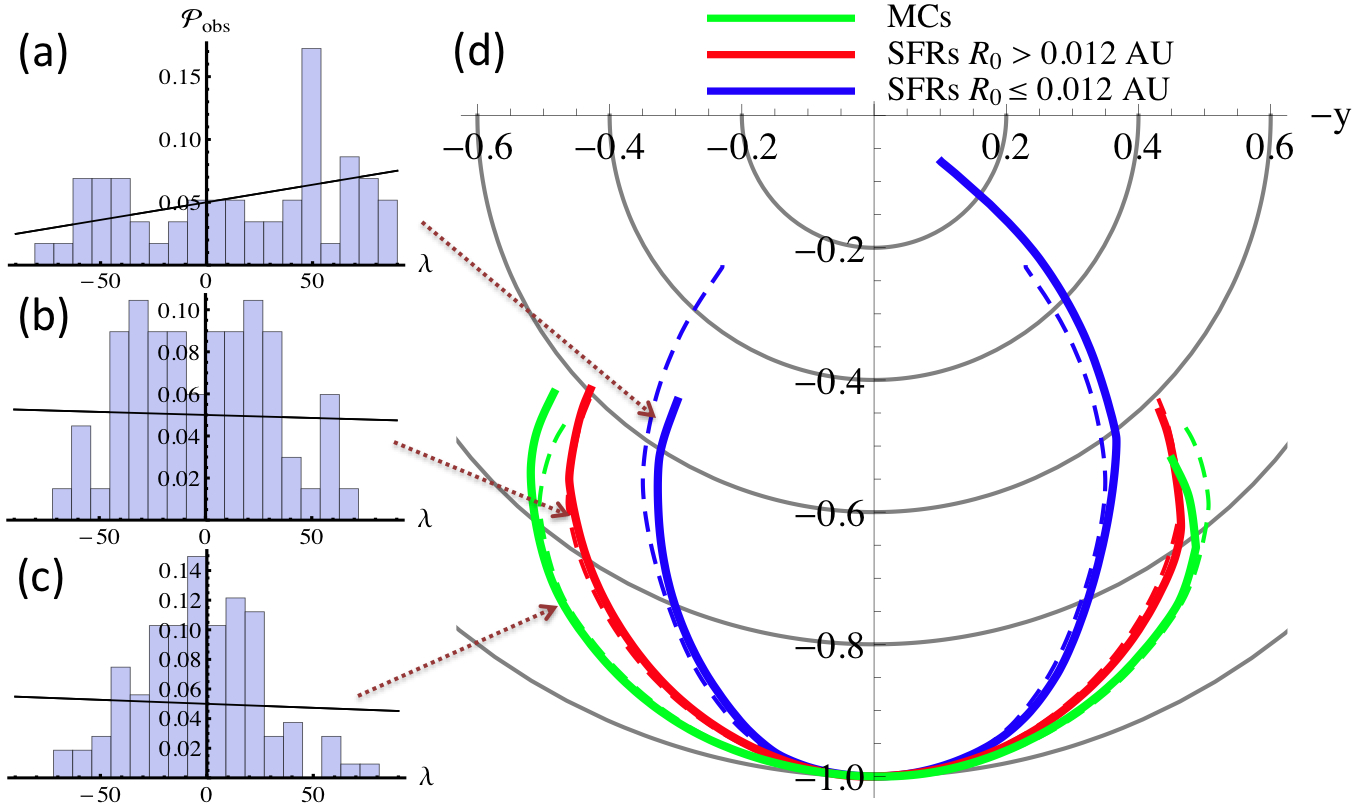} }
\caption{Probability distributions $\pobsl$ for (a) very {small, (b) larger small FRs, and (c)} MCs.  (d) Mean FR axis deduced from a spline interpolation of $\pobsl$ shown in panels a-c for $\phimax = 45 \degree$.  This relies on the plausible hypothesis that FRs have a comparable mean structure and that the spacecraft has an equi-probability of crossing {along} the FRs projected on the 1~AU sphere. The axis shapes with symmetrised $\pobsl$ are shown with thinner dashed lines.}
 \label{fig:shape_axis}
\end{figure}  

The continuous distribution $\pobs (\lA')$ present in \eqs{varphi}{rho} can be derived from observations by Hermite or spline polynome interpolation, or by a fit (such as a straight line) of the distributions shown in \figsss{shape_axis}a-c.  The results for small FRs are weakly sensitive to the method selected, similarly to MCs, because the shape is obtained via an integration of the data \citep[see][]{Janvier13}.   

{Since small FRs are not imaged by heliospheric imagers, we have selected the same $\phimax $ for all the FRs} (selected large for a better viewing). We found that the FR axis has significantly different shapes when it is associated with the three different distributions shown in \figsss{shape_axis}a-c. In particular, the very small FRs have the broadest $\pobsl$ distribution, implying a more bent axis  {(\fig{shape_axis}d).}  They also have the most asymmetric distribution, as globally outlined by the linear fit (\fig{shape_axis}a).  This implies an asymmetry of the FR axis (see blue line in \fig{shape_axis}d). This asymmetry is outlined by the comparison with the FR axis shape computed if we were to consider a symmetric $\pobsl$ (dashed line).  However, the asymmetry of the legs is weak and far from the Parker's spiral shape of the SW magnetic field.  

The distribution $\pobsl$ is both narrower and more symmetric for larger FR radius, implying a flatter and more symmetric FR shape.   For such FRs, there is no evidence of a leg deformation by the Sun rotation, at least for the front part of the legs (for the parts closer to the Sun the spacecraft trajectory has a too small angle with the FR axis to allow a fit with a straight FR model, or even to recognize that a FR was crossed \citep{Owens12}).  Finally, changing the constant $\phimax $ only rescales all the FR shapes in the $\rho$ and $\varphi$ directions according to \eqs{varphi}{rho} (see examples Figure 10 of \citet{Janvier13}).

\section{Conclusions} 
      \label{sect:Conclusion} 
 
The main aim of this study is to compare the physical properties of small flux ropes (small FRs) with that of magnetic clouds (MCs).    For that, we analysed the statistical properties of two lists of FRs, one of small FRs \citep{Feng08} and one of MCs \citep{Lepping10}. Whether the small FR population is only a smaller size version of MCs is a difficult question to answer since those two populations have both common and different properties. Are the common properties mostly a consequence of the same physics applied to FRs embedded in a common solar wind (SW) environment, or are they a signature of a common formation process? Are the different properties mostly a scale effect or are they a signature of different  formation processes?  
We summarize below the results of the present study, as well as the previous ones, with the aim to answer these questions.

We find that small FRs and MCs have {\bf intrinsic similarities} such as a larger field strength and a larger outward velocity with an increasing FR radius (\fig{V(R)_Bo(R)}).  These positive correlations are plausibly the remnant traces of their solar launch conditions.  They also both have a statistically larger radius at their apex than in their legs while the measurements are done in both cases at 1 AU (\fig{R_apex_legs}).  These similarities add up to the previously found ones (see Introduction), such as {their FR structure and} the absence of a special structure in the plasma density across the FR.

There are also {\bf similarities} between small FR and MC properties that are direct consequences of the interaction with the surrounding SW. A first similarity is the presence of a sheath of denser plasma at their front (weaker for small FRs as they are slower). A second one comes from the total pressure balance between the FR and its surroundings. This implies that both small FRs and MCs are expanding as they move away from the Sun in a progressively lower pressure environment.  However, the normalized expansion rate [$\zeta $, \eq{zeta}] for small FRs is half that of MCs.  We interpret this lower expansion rate as a consequence of a lower total pressure decrease with solar distance in the surrounding SW, as most small FRs travel closely to the heliospheric current sheet. Also, due to their small sizes, small FRs are mostly affected by this local SW pressure.  Finally, the expansion of most small FRs is not detectable by \insitu\ observations because of their small sizes.  {It implies a small expansion velocity that is covered by the presence of} velocity fluctuations (\fig{DV}).  

Several {\bf differences} between small FRs and MCs are linked to their field strength difference. The small FRs have a weaker axial magnetic flux, and consequently a lower  
magnetic tension. As such, they are more easily deformed/rotated than MCs.   As a consequence, the orientation of very small FRs tend to be parallel to the ecliptic, within two magnetic sectors, while the orientation of MC axis direction has a more isotropic distribution (\fig{slope_Pi}).    Next, both small FRs and MCs have a lower plasma $\beta$ than in the surrounding SW but the effect is much weaker in small FRs as, first, the field strength is lower, and second, the temperature and density have no significant contrast with the surrounding SW.  
 
There are also a few {\bf intrinsic differences} between small FRs and MCs.  The most important is that small FRs and MCs have two different distributions of radius: a power-law for small FRs and a Gaussian-like for MCs \citep{Janvier14}.
Another difference is the absence of a lower proton temperature in small FRs, while it is one criterium used to define MCs.  This temperature difference could simply be due to a lower expansion rate for small FRs (factor two lower on $\zeta $), or a size effect (such as the high energy particles more likely to enter small FRs than MCs), or due to the intrinsic property of different formation processes.  
   
{The results described above, when considered altogether, show that the properties of small FR and MC properties are closely related. Some of the properties, that appear different, could be explained by a difference in the size and the field strength of these structures. Also, small FRs are more affected than MCs by the surrounding SW, although it} is also possible that all FR properties are modified by the same interplanetary 
physical processes, so that their intrinsic source properties mostly disappear as they reach 1~AU. {In order to better understand the main solar processes that generate small flux ropes on the one hand, and MCs on the other hand, it would be important to remove as much as possible these propagation effects.}  Presently, we can still argue that the distribution of sizes is difficult to explain with the same FR origin, so small FRs are not just the continuation to smaller scales of MCs \citep{Janvier14}.

 Finally, we studied the probability distributions [$\pobsAi,\pobsl$] of the FR axis orientations.  In particular, the distribution of the location angle [$\lA$, defined in \fig{schema}] is directly related to the global shape of the FR axis.  We find that $\pobsl$ is only affected by the FR radius.   $\pobsl$ is nearly flat for very small FRs ($R \le 0.01$~AU), and comparable to MC distribution for the larger small FRs (\fig{slope_Plambda}).  Supposing that small FRs have a coherent global FR shape, as MCs, this implies that very small FRs have a more bent FR axis than larger FRs for the same angular extension $\phimax$ (\fig{shape_axis}).   Only a weak asymmetry is found between the eastern and the western legs, so FRs are not significantly deformed in a Parker-like spiral, {at least in a broad range around their apex}. Such shape determination has implications on estimations of global quantities, such as magnetic helicity.  These will be explored in a next study. 

{Finally, a significant advance on the subject of the nature of flux ropes in the interplanetary medium is expected to be provided by the future Solar Orbiter and Solar Probe Plus missions. Indeed, with these missions,} \insitu\ data will be collected much closer to the Sun, {where the propagation effects are weaker than at 1 AU.}

\begin{acknowledgements}
The data used in the present paper are provided by Feng et al. \cite[see][]{Feng08} for small flux ropes, and Lepping and Wu  for magnetic clouds \cite[also at \url{http://wind.nasa.gov/mfi/mag_cloud_S1.html}, see][]{Lepping10}. We thanks these authors for making those data publicly available.
This work was partially supported by  supported by the Argentinean grant UBACyT 20020120100220 and PICT-2013-1462 (SD), and by a one month invitation of SD by Paris Observatory.
SD is member of the Carrera del Investigador Cien\-t\'\i fi\-co, CONICET.
\end{acknowledgements}

   
\bibliographystyle{agufull08}

\bibliography{properties_SFR-revised-final-wobold}  

\IfFileExists{\jobname.bbl}{} {\typeout{}
\typeout{****************************************************}
\typeout{****************************************************}
\typeout{** Please run "bibtex \jobname" to obtain} \typeout{**
the bibliography and then re-run LaTeX} \typeout{** twice to fix
the references !}
\typeout{****************************************************}
\typeout{****************************************************}
\typeout{}}

\end{article}
\end{document}